\documentclass[12pt,a4paper]{article}
\usepackage{textcomp}
\usepackage{listings}
\usepackage[space=true]{accsupp}
\newcommand{\copyablespace}{\BeginAccSupp{method=hex,unicode,ActualText=00A0}\EndAccSupp{}}
\usepackage[T1]{fontenc}
\usepackage[english]{babel}
\usepackage{hyperref}
\usepackage{graphicx}
\usepackage{amsmath}
\usepackage{amssymb}
\usepackage[table]{xcolor}
\usepackage{fancyhdr}
\usepackage{longtable}
\usepackage{multicol}
\usepackage{enumerate}
\usepackage{enumitem}
\usepackage{multirow}
\usepackage{natbib}

\newtheorem{definition}{Definition}
\newtheorem{example}{Example}

\title{Filling in pattern designs for incomplete pairwise comparison matrices: (quasi-)regular graphs with minimal diameter}
\author{S\'andor Boz\'oki$^{1,2}$, Zsombor Sz\'adoczki$^{1,2}$, Hailemariam Abebe Tekile$^{3}$}
\begin{document}

%\pagenumbering{gobble}
\pagenumbering{arabic}

\maketitle
\begin{center}
$^{1}$ Research Laboratory on Engineering \& Management Intelligence \\
Institute for Computer Science and Control (SZTAKI) \\
Budapest, Hungary\\
$^{2}$ Department of Operations Research and Actuarial Sciences \\
Corvinus University of Budapest \\
$^{3}$ Department of Industrial Engineering \\
University of Trento, Italy 

\end{center}

\vspace{2cm}

\begin{abstract}

\noindent
Multicriteria Decision Making problems are important both for individuals and groups. 
Pairwise comparisons have become popular in the theory and practice of preference modelling and quantification.
We focus on decision problems where the set of pairwise comparisons can be chosen, i.e., it is not given
a priori.
The objective of this paper is to provide recommendations for filling patterns of incomplete pairwise comparison matrices (PCMs) 
based on their graph representation. 
Regularity means that each item is compared to others for the same number of times, resulting in a kind of symmetry.
A graph on an odd number of vertices is called quasi-regular, if the degree of every vertex is the same odd number, except for one vertex whose degree is larger by one.
If there is a pair of items such that their shortest connecting path is very long,
the comparison between these two items relies on many intermediate comparisons,
and is possibly biased by all of their errors. 
Such an example was found in \cite{Tekile} where
the graph generated from the table tennis players' matches included a long shortest path between two vertices (players), and the calculated result appeared to be misleading.

If the diameter of the graph of comparisons is low as possible (among the graphs of the same number of edges),
we can avoid, or, at least decrease, such cumulated errors.

The aim of our research is to find graphs, 
among regular and quasi-regular ones, with minimal diameter. 
Both theorists and practitioners can use the results, given in several formats in the appendix: graph, adjacency matrix, list of edges. 

\end{abstract}

\renewcommand{\baselinestretch}{1.24} \normalsize

\section{Introduction}

Multicriteria Decision Making is a really important tool both at an individual and at an organizational level. We can almost think about any kind of ranking of alternatives or weighting of criteria, like tenders, selection among schools or job offers, selection among the evaluation of different projects in an enterprise, etc.

One of the most commonly used technique in connection with the Multicriteria Decision Making is the method of the pairwise comparison matrices \citep{Saaty}. One can apply this technique both for determining the weights of the different criteria and for the rating of the alternatives according to a criterion. Usually we denote the number of criteria or alternatives by $n$, which means the pairwise comparison matrix is an $n\times n$ matrix often denoted by $A$. In this case the $ij$-th element of the $A$ matrix, $a_{ij}$ shows how many times the $i$-th item is larger/better than the $j$-th element.

Formally, matrix $A$ is called a pairwise comparison matrix (PCM) if it is positive ($a_{ij}>0$ for $\forall$  $ i $  and  $ j$) and reciprocal ($1/a_{ij}  = a_{ji}$ for $\forall$ $ i $ and $ j$) \citep{Saaty}, which also indicates that $a_{ii}=1$ for  $\forall$ $i$.

When some elements of a PCM are missing we call it an incomplete PCM. There could be many different reasons why these elements are absent, some data could have been lost or the comparisons are simply not possible (for instance in sports \citep{BozokiCsatoTemesi}).

The most interesting case for us is when the decision makers do not have time, willingness or the possibility 
to make all the $n(n-1)/2$ comparisons. 

In this article we would like to show which comparisons are the most important ones to be made, or more precisely what pattern of comparisons are recommended to be made in order to get the best approximation of the decision makers' original preferences in different cases, when we have some assumptions on our Multicriteria Decision Making (MCDM) problem. The graph representation of the pairwise comparisons is a natural and convenient tool to examine our question, thus we will use this throughout the paper.

Special structures from incomplete pairwise comparison matrices include
(i) spanning tree, in particular if one row/column is filled in completely (its associated graph is the star graph)
(ii) two rows/columns are filled in completely (its associated graph is the union of two star graphs) \cite{Rezaei2015}
(iii) more or less regular graphs, for example in case of sport competitions \cite{Csato2013}, where the number
of matches played equals for every player or team, at least in the first phase (before the knockout phase).

These examples do not take the diameter into consideration, and the first two examples lack regularity, too.
Regularity means that each item is compared to others for the same number of times (if the cardinality of the
items to compare is odd, one of the degrees can be smaller or greater - in our analysis, greater - by one),
resulting in a kind of symmetry.
In the set of connected graphs, diameter can be considered as a measure of closeness, or a stronger type
of connectedness.  

To understand the used methodology we have to define some basic mathematical concepts, which takes place in the second section of the article. Later on we assume that we know the number $n$ of alternatives or criteria, it is also a key assumption through our paper that the graph that representing the MCDM problem is $k$-regular and we also know (or with the help of the other inputs we can determine) the diameter $d$ of the graph. In the third section we provide a systematic collection of suggested incomplete pairwise comparisons' patterns with the help of the above-mentioned inputs and the/some graphs for the examined cases. In Section 4 we make our conclusion and provide further research questions closely connected to the discussed topic.

Note that regular graphs can have large diameter, e.g., a cycle on $n$ vertices is 2-regular and has diameter
$d= \lfloor n/2 \rfloor$. The star graph, mentioned among the examples, has minimal diameter 2, but it is
far from being regular. Our aim is to find the graphs, among (quasi-)regular ones, with minimal diameter. We are
especially interested in the smallest nontrivial values of the diameter, namely $d= 2$ and $d=3$.

\section{Basic concepts of the graph representation}

The graph representation of paired comparisons has already been used in the 1940s \citep{KendallSmith}. Of course after the widespread application of PCMs and incomplete PCMs it has become a really common method in the literature, see for instance \cite{Blanquero}, \cite{Csato} or \cite{Gass}.

Usually in these articles the authors use directed graphs for the representation, because they distinguish the preferred item from the less preferred one in every pair. In our approach the only important thing is the following: is there a comparison between the two elements or not. This means that we use undirected graphs, where the vertices denote the criteria or the alternatives. There is an edge between two vertices only if the decision makers made their comparison for the two respective items, while there is not an edge between two vertices only if the decision makers have not made their comparison for the two units (the respective element of the PCM is missing). In order to understand the concepts so far, there is a small example below:

\begin{example}
\label{example:1}
Let us assume that there are 4 criteria ($n=4$) and our decision maker already answered some questions, denoted their locations in the matrix by $\bullet$ and their reciprocal values by $\circ$, which lead to the following incomplete PCM:

$$A = \begin{bmatrix}
1       & \bullet & \bullet &          \\
\circ &     1     &         & \bullet  \\
\circ &           &      1  & \bullet  \\
        &  \circ  &  \circ  &     1
\end{bmatrix}$$

\bigskip
\noindent
This incomplete PCM is represented by the graph in Figure~\ref{fig:1}.

\begin{figure}[ht!]
	\centering
	\includegraphics[width=0.4\textwidth, height=0.25\textheight]{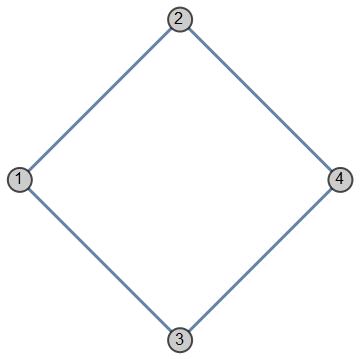}
	\label{fig:1}
	\caption{Graph representation example}
\end{figure}

\noindent As we can see there is no edge between the first and the fourth vertices, where the PCM has missing values and there is no edge between the second and third vertices, where the situation is the same. 
There is an edge between every other pair, where we have no missing values in the PCM.  

\end{example}

We assume that the representing graphs are connected and $k$-regular through our paper, thus we need some definitions to make these concepts clear.

\begin{definition}[Connected graph]

In an undirected graph, two vertices $u$ and $v$ are called connected if the graph contains a path from $u$ to $v$. A graph is said to be connected if every pair of vertices in the graph is connected.

\end{definition}

\begin{definition}[$k$-regular graph]

A graph is called $k$-regular if every vertex has $k$ neighbours, which means that the degree of every vertex is $k$.

\end{definition}

\begin{definition}[$k$-quasi-regular graph]

A graph is called $k$-quasi-regular if exactly one vertex has degree $k+1$, and all the other vertices have degree $k$.

\end{definition}

The $k$-regularity basically means that the vertices are not distinguished, there is no particular vertex as, for example, in the case of the star graph, thus we would like to avoid the cases when the elimination of relatively few vertices would lead to the disintegration of the whole comparison system \citep{Tekile}. While the connectedness is really important, because to approximate the decision makers' preferences well, we need to have at least indirect comparisons between the different criteria, otherwise we cannot say anything about the relation between certain elements.

However, it is also notable that we would like to avoid the cases when two items are compared only indirectly through a very long path, because this could aggregate the small, tolerable errors of the different comparisons and we could end up with an intolerably large error in the relation between the two elements. To measure this problem we can use the diameter of the representing graph:

\begin{definition}[The diameter of a graph]

The diameter (denoted by $d$) of a graph G is the length of the longest shortest path between any two vertices:

$$d=\max_{u,v \in V(G)}\ell(u,v),$$

\noindent
where $V(G)$ denotes the set of vertices of $G$ and $\ell(.,.)$ is the graph distance between two vertices, namely the length of the shortest path between them.

\end{definition}

Briefly from now on we will examine graphs representing MCDM problems defined by the following inputs: $(n,k,d)$, where $n$ is the number of vertices (criteria), $k$ shows the level of regularity of the graph and $d$ is the diameter of the graph.

\section{Results}

First of all it is a key step to determine which cases are interesting for us considering our inputs. It is important to emphasize that we deal with unlabelled graphs, because we are trying to find out what kind of patterns are needed in the comparisons for different instances, thus  if we exchange the 'names' of two criteria (like if we would change '1' and '2' in Example~\ref{example:1}) the pattern would be the same.

Then we can consider the regularity parameter, $k$. The $k=1$ case is possible only when $n$ is even, but they are not connected except for $n=2$, so this is not really interesting for us. When $k=2$ there is only one connected graph for every $n$, namely the cycle, for which $d=\lfloor n/2 \rfloor$ as already mentioned in the introduction.

The larger regularity parameters could be interesting for us, but of course we need a reasonable upper bound for the number of criteria, $n$, which is also an indirect upper bound for $k$. In our research we examined the $n=3,4,\ldots,20$ cases, because in the one hand for larger $n$ parameters, some computations become really difficult, and in the other hand we think that it is reasonable to assume that we do not have more than 20 important criteria or 20 relatively good alternatives in most of the MCDM problems.

The smaller the $d$ parameter is, the more stable or the more trustworthy our system of comparisons is. This means that in an optimal case we would like to minimize this parameter, while the number of the criteria ($n$) is always a given exogenous parameter in our MCDM problems. As we mentioned above, $k$ is crucial to avoid the cases when some criteria (vertices) would be too important in the system, however it also shows us how many comparisons have to be made, because every vertex has a degree of $k$, which means the number of edges is $nk/2$. Thus if our decision makers would like to spend the shortest time with the creation of the PCM, we should choose a small $k$ parameter. But, of course, as usually happens in these situations, there is a trade off between the parameters, because for many criteria (large $n$) the smaller regularity ($k$) will cause a bigger diameter ($d$), namely a more fragile system of comparisons.

In this paper we would like to provide a list of graphs which shows the patterns of the comparisons that have to be made in case of different parameters. We used an algorithm which defines the graph(s) with the smallest diameter ($d$ parameter) for a given $(n,k)$ pair. With the help of these results it was easy to determine which $k$ is the smallest that is needed to reach a given $d$ for a given $n$. We found that, with the chosen upper bound of $n$ (20) the interesting values for the regularity are $k=3,4,5$, while the interesting values for the diameter of the graph are $d=2,3$. For a general MCDM problem probably instead of $k$, it would give more information if we considered an indicator that shows how far we are from the 'extreme' case when the decision makers have to make all the comparisons. This would mean $n(n-1)/2$ comparisons instead of our $nk/2$ in case of regular graphs or $(nk+1)/2$ in case of quasi-regular graphs, therefore the completion ratio is defined as follows:
\[
c= \left\{ 
\begin{tabular}{ll} 
$\frac{nk/2}{n(n-1)/2}$ & \text{if $n$ or $k$ is even} \\
$\frac{(nk+1)/2}{n(n-1)/2}$ & \text{if $n$ and $k$ are odd} 
\end{tabular} 
\right.
\]
that we will calculate for every instance.

Now we will present the results of the algorithm which gives the graphs with the smallest diameter for a given $(n,k)$ pair. Of course $d=1$ would mean a complete graph that is not reachable for many $(n,k)$ pairs, and also not so interesting for us, thus Table~\hyperlink{tab:1}{1} shows the cases when $k=3$ and $d=2$ is the minimal value of the parameter. It is also important to note that $k=3$ is only possible when $n$ is even, but when it is odd, we examine graphs where all vertices' degrees are 3 except one where it is 4, because these are the closest to 3-regularity.

\begin{table}[ht!]
\centering
\renewcommand{\baselinestretch}{0.75} 
\normalsize
\begin{tabular}{|m{0.9cm}|m{2.5cm}|m{3.5cm}|m{0.9cm}|m{2.5cm}|m{3.5cm}|}
	\hline\begin{center}$\boldsymbol{k}$\textbf{=3} \end{center} & \begin{center} \hypertarget{tab:1}{Graph} \end{center} &  \begin{center} Further information \end{center} & \begin{center}	$\boldsymbol{k}$\textbf{=3} \end{center} & \begin{center} Graph \end{center} &  \begin{center} Further information \end{center}  \\ \hline \hline

\begin{center} $\boldsymbol{n}$\textbf{=5} \end{center} &\text{ }
	\includegraphics[width=0.15\textwidth,  height=0.07\paperheight]{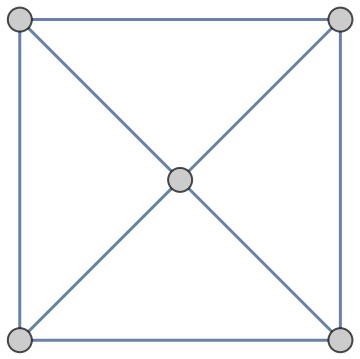}  \begin{center}  \end{center} & \begin{itemize} \item 8/10 comparisons ($c=0.8$)  \item $\geq 2$ graphs\end{itemize}  &\begin{center} $\boldsymbol{n}$\textbf{=6} \end{center} & \text{ }
	\includegraphics[width=0.15\textwidth,  height=0.07\paperheight]{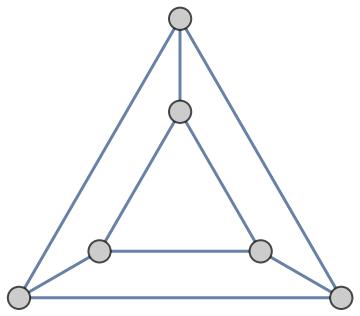}  \begin{center} 3-prism graph \end{center} & \begin{itemize} \item 9/15 comparisons ($c=0.6$)  \item 2 graphs \item The other solution is the bipartite graph $K_{3,3}$ \end{itemize} \\ \hline 

	\begin{center} $\boldsymbol{n}$\textbf{=7} \end{center} &\text{ }
	\includegraphics[width=0.15\textwidth,  height=0.07\paperheight]{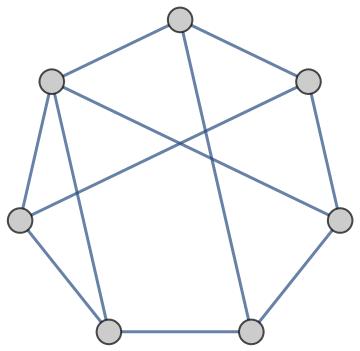}  \begin{center}  \end{center} &\begin{itemize} \item 11/21 comparisons ($c\approx0.524$)  \item $\geq 2$ graphs \end{itemize} &\begin{center} $\boldsymbol{n}$\textbf{=8}\end{center} & \text{ }
	\includegraphics[width=0.15\textwidth,  height=0.07\paperheight]{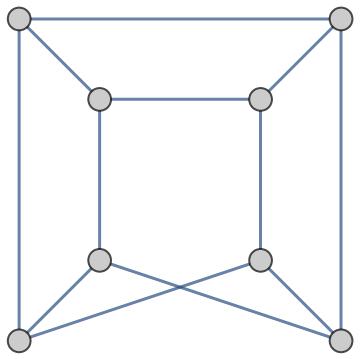}  \begin{center} Wagner graph \end{center} & \begin{itemize} \item 12/28 comparisons ($c\approx0.429$)  \item 2 graphs \end{itemize} \\ \hline 
	
\begin{center} $\boldsymbol{n}$\textbf{=9} \end{center}  &\text{ }
	\includegraphics[width=0.15\textwidth,  height=0.07\paperheight]{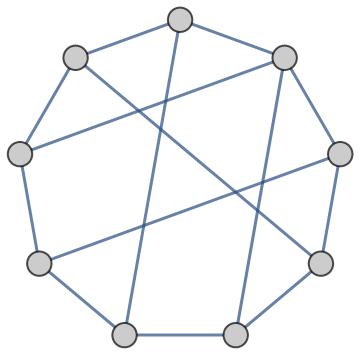}  \begin{center}  \end{center} &\begin{itemize} \item 14/36 comparisons ($c\approx0.389$)  \item $\geq 1$ graph \end{itemize} &\begin{center} $\boldsymbol{n}$\textbf{=10} \end{center} & \text{ }
    \includegraphics[width=0.15\textwidth,  height=0.07\paperheight]{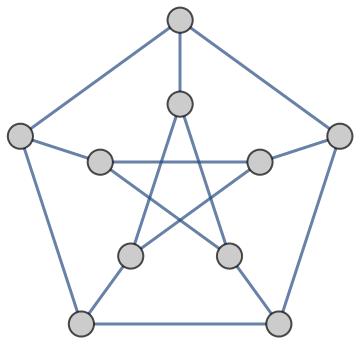}  \begin{center} Petersen graph \end{center} & \begin{itemize} \item 15/45 comparisons ($c\approx0.333$)  \item Unique graph \end{itemize}\\ \hline

	\end{tabular}
	\renewcommand{\baselinestretch}{1} 
	\Large
	\caption{$k=3$-(quasi-)regular graphs on $n$ vertices with minimal diameter $d=2$}
	\end{table}
	
\renewcommand{\baselinestretch}{1.24} 

We can see that with $k=3$ the minimal diameter can be 2 until we have 10 vertices. Of course for $n\leq3$ the 3-regularity is not possible, and for $n=4$ the diameter is 1, because this is a complete graph, but those are really simple (trivial) cases with few possibilities, that is why we skip those in the table. It is also notable that the completion ratio ($c$) even reach $1/3$ when we have 10 vertices (it is obviously decreasing in $n$). And we should emphasize the fact that there are only a few graphs for every $(n,k)$ pair with the minimal diameter, and one of them is often a bipartite graph that is not the best design in a decision problem, because the two groups are always compared through the other ones \citep{Csato}. Where the table contains '$\geq\ldots$ graphs' that means we have not checked all the possible cases with minimal diameter, but in connection with decision making problems it is enough to see that there is one pattern that satisfies the needed properties.

If we go on to larger graphs ($n>10$), then we will find  that the smallest reachable diameter changes to $d=3$, but it is also true that at first we have so many graphs that satisfies these properties. However as we examine the $n=18$ or the $n=20$ cases, we can see that there is only one graph that fulfils our assumptions \citep{Pratt}. The results in case of larger graphs, with 3-regularity and 3 as the minimal diameter can be found in Table \hyperlink{tab:2}{2}.

	\begin{table}[ht!]
	\centering
	\renewcommand{\baselinestretch}{0.75}
	\small
	\begin{tabular}{|m{0.9cm}|m{2.8cm}|m{3cm}|m{0.9cm}|m{2.8cm}|m{3cm}|}
	\hline \begin{center}	$\boldsymbol{k}$\textbf{=3} \end{center} & \begin{center} \hypertarget{tab:2}{Graph} \end{center} &  \begin{center} Further information \end{center} & \begin{center}	$\boldsymbol{k}$\textbf{=3} \end{center} & \begin{center} Graph \end{center} &  \begin{center} Further information \end{center} \\ \hline \hline

	\begin{center} $\boldsymbol{n}$\textbf{=11} \end{center} & \text{ }
	\includegraphics[width=0.13\textwidth,  height=0.05\paperheight]{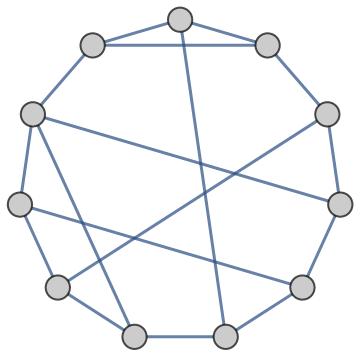}  \begin{center}  \end{center} & \begin{itemize} \item 17/55 comparisons ($c\approx0.309$)  \item $\geq34$ graphs \end{itemize} & \begin{center} $\boldsymbol{n}$\textbf{=12} \end{center} & \text{ }
	\includegraphics[width=0.13\textwidth,  height=0.05\paperheight]{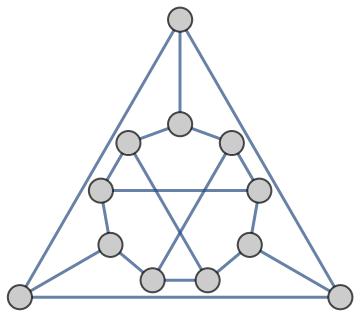}  \begin{center} Tietze graph \end{center} & \begin{itemize} \item 18/66 comparisons ($c\approx0.273$)  \item 34 graphs \end{itemize} \\ \hline 
	
\begin{center} $\boldsymbol{n}$\textbf{=13} \end{center} &
	\text{ } \includegraphics[width=0.13\textwidth,  height=0.05\paperheight]{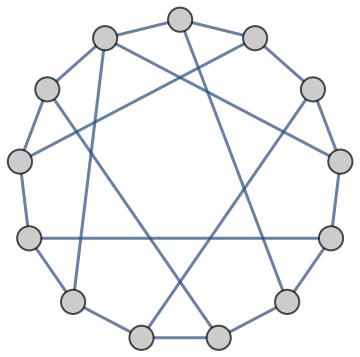}  \begin{center}  \end{center} & \begin{itemize} \item 20/78 comparisons ($c\approx0.256$)  \item $\geq 34$ graphs \end{itemize} &
	\begin{center} $\boldsymbol{n}$\textbf{=14} \end{center} &
	\text{ } \includegraphics[width=0.13\textwidth,  height=0.05\paperheight]{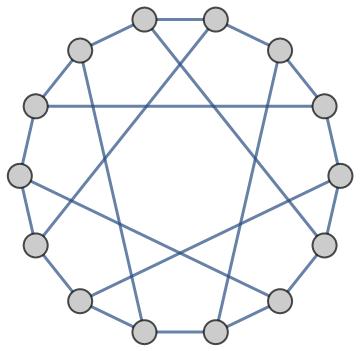}  \begin{center} Heawood graph \end{center} & \begin{itemize} \item 21/91 comparisons ($c\approx0.231$)  \item 34 graphs \end{itemize} \\ \hline 

	\begin{center}	$\boldsymbol{n}$\textbf{=15} \end{center} & \text{ } \includegraphics[width=0.13\textwidth,  height=0.05\paperheight]{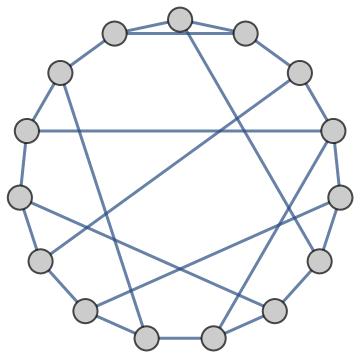}  \begin{center}  \end{center} & \begin{itemize} \item 23/105 comparisons ($c=0.219$)  \item $\geq14$ graphs \end{itemize} &
	\begin{center}	$\boldsymbol{n}$\textbf{=16} \end{center} & \text{ } \includegraphics[width=0.13\textwidth,  height=0.05\paperheight]{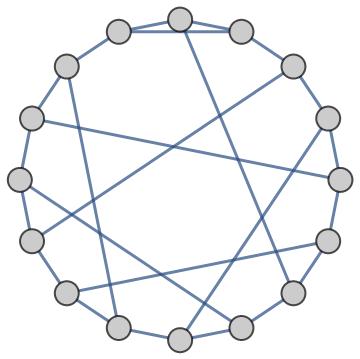}  \begin{center}  \end{center} & \begin{itemize} \item 24/120 comparisons ($c=0.2$)  \item  14 graphs \end{itemize} \\ \hline

	\begin{center}	$\boldsymbol{n}$\textbf{=17} \end{center} & \text{ } \includegraphics[width=0.13\textwidth,  height=0.05\paperheight]{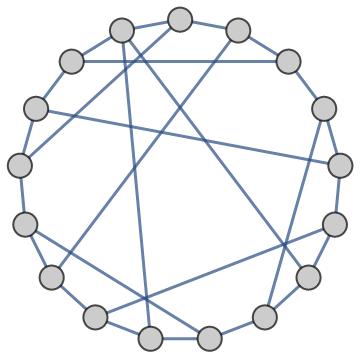}  \begin{center}  \end{center} & \begin{itemize} \item 26/136 comparisons ($c\approx0.191$)  \item  $\geq1$ graph \end{itemize} &
	\begin{center}	$\boldsymbol{n}$\textbf{=18} \end{center} & \text{ } \includegraphics[width=0.13\textwidth,  height=0.05\paperheight]{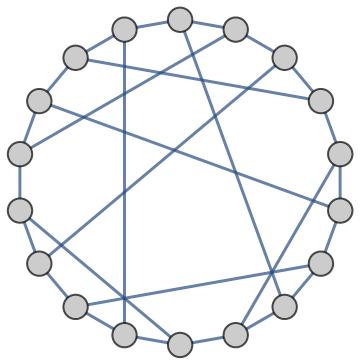}  \begin{center} (3,3) graph on 18 vertices \end{center} & \begin{itemize} \item 27/153 comparisons ($c\approx0.176$)  \item  Unique graph \end{itemize} \\ \hline

	\begin{center}	$\boldsymbol{n}$\textbf{=19} \end{center}& \text{ } \includegraphics[width=0.13\textwidth,  height=0.05\paperheight]{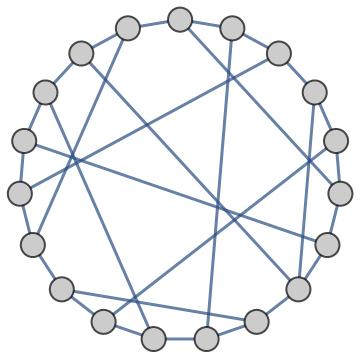}  \begin{center}   \end{center} & \begin{itemize} \item 29/171 comparisons ($c\approx0.170$ \item $\geq1$ graph \end{itemize} &
	\begin{center}	$\boldsymbol{n}$\textbf{=20} \end{center}& \text{ } \includegraphics[width=0.13\textwidth,  height=0.05\paperheight]{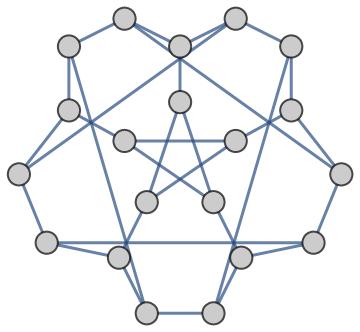}  \begin{center} \footnotesize{(3,3)-graph on 20 vertices (C5xF4)} \end{center} & \begin{itemize} \item 30/190 comparisons ($c\approx0.158$ \item Unique graph \end{itemize} \\ \hline
	\end{tabular}
	
	\renewcommand{\baselinestretch}{1}
	\Large
	\caption{$k=3$-(quasi-)regular graphs on $n$ vertices with minimal diameter $d=3$}
	\end{table}

\renewcommand{\baselinestretch}{1.24}	

As we can see the completion ratio is still decreasing in $n$ and on larger graphs it can be taken below 0.2. It is also true that we still do not need to answer for more than 30 questions for an MCDM problem with 20 criteria, which can be really useful.

We discussed all the possible cases for $k=3$ and $n\leq20$, and we saw that the minimal diameter is 2 or 3 here. We also mentioned that the 1 diameter would mean a complete graph and a complete PCM, so that is not interesting for us. This means that if we would like to examine the graphs where $k=4$ it is obvious that the minimal diameter would be 2 until $n=10$, but it is not so important to make so many comparisons because this property can be reached with $k=3$, too. Thus for $k=4$ the interesting cases start above 10 vertices, and the question is that can we reach a smaller diameter (a more stable system of comparisons) with the rise of the answered questions. We found that with $k=4$ we can get 2 as the minimal diameter until $n=15$, but for larger values of $n$, it will be 3 again which can be also reached by $k=3$, thus we would not recommend these combinations of parameters. The results for $(11\leq n\leq15,k=4)$ are shown in Table \hyperlink{tab:3}{3}. It is also important to note that $k=4$ is possible in case of both odd and even values of $n$, thus now we do not have to pay special attention to this.

	\begin{table}[ht!]
	\centering
	\renewcommand{\baselinestretch}{0.75}
	\small
	\begin{tabular}{|m{1cm}|m{3cm}|m{5cm}|}
	\hline \begin{center}	$\boldsymbol{k}$\textbf{=4} \end{center} & \begin{center} \hypertarget{tab:3}{Graph} \end{center} & \begin{center} Further information \end{center} \\ \hline \hline

	\begin{center} $\boldsymbol{n}$\textbf{=11} \end{center} & \text{ }
	\raisebox{-\totalheight}{\includegraphics[width=0.15\textwidth,  height=0.05\paperheight]{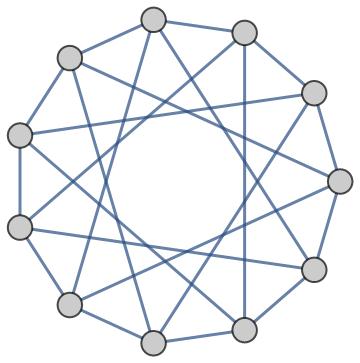}}  \begin{center} \footnotesize{4-Andr\'asfai graph} \end{center}& \begin{itemize} \item 22/55 comparisons ($c=0.4$)  \item 37 graphs \end{itemize} \\ \hline
	
	\begin{center} $\boldsymbol{n}$\textbf{=12} \end{center} & \text{ }
	\raisebox{-\totalheight}{\includegraphics[width=0.15\textwidth,  height=0.05\paperheight]{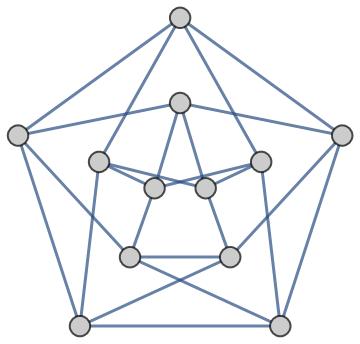}}  \begin{center} \footnotesize{Chv\'atal graph} \end{center}& \begin{itemize} \item 24/66 comparisons ($c\approx0.364$)  \item 26 graphs \end{itemize} \\ \hline

	\begin{center} $\boldsymbol{n}$\textbf{=13} \end{center} & \text{ }
	\raisebox{-\totalheight}{\includegraphics[width=0.15\textwidth,  height=0.05\paperheight]{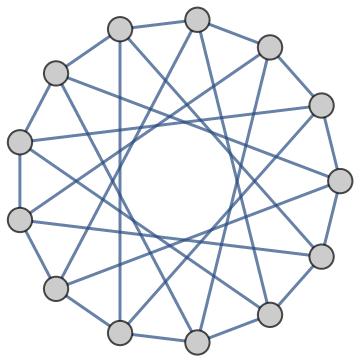}}  \begin{center} \footnotesize{13-cyclotomic graph} \end{center} & \begin{itemize} \item 26/78 comparisons ($c\approx0.333$)  \item 10 graphs \end{itemize}\\ \hline 

	\begin{center} $\boldsymbol{n}$\textbf{=14} \end{center} & \text{ }
	\raisebox{-\totalheight}{\includegraphics[width=0.15\textwidth,  height=0.05\paperheight]{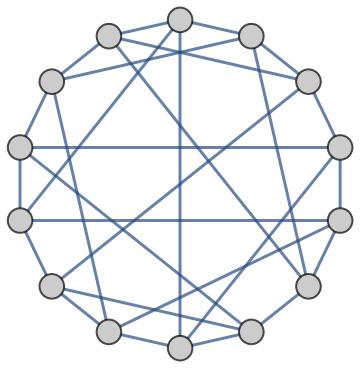}}  \begin{center} \footnotesize{Unique graph on 14 vertices} \end{center}& \begin{itemize} \item 28/91 comparisons ($c\approx0.308$)  \item Unique graph \end{itemize}\\ \hline 

	\begin{center} $\boldsymbol{n}$\textbf{=15} \end{center} & \text{ }
	\raisebox{-\totalheight}{\includegraphics[width=0.15\textwidth,  height=0.05\paperheight]{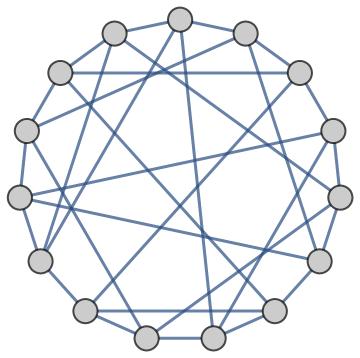}}  \begin{center} \footnotesize{Unique graph on 15 vertices} \end{center}& \begin{itemize} \item 30/105 comparisons ($c\approx0.286$)  \item Unique graph \end{itemize}\\ \hline 
	\end{tabular}
	
	\renewcommand{\baselinestretch}{1}
	\Large
	\caption{$k=4$-regular graphs on $n$ vertices with minimal diameter $d=2$}
	\end{table}
	
	\renewcommand{\baselinestretch}{1.24}
	
As we can see, the completion ratio is increasing in $k$, so we cannot get so small $c$ values as in the former table, however the system of comparisons will be more stable even on many vertices, because the smallest diameter is 2 here. It is also really interesting that, for larger graphs and regularity levels, the number of connected graphs increase very rapidly. For instance, when we have 15 vertices, there are 805 491 connected 4-regular graphs (that means 805 491 possible filling patterns of the PCM), and only one has 2 as its diameter. Our algorithms and methodology has a strong relationship with the so-called degree-diameter problem that is well known in the literature of mathematics (\cite{DinneenHafner}, \cite{EyalSiran}), but they are looking for the largest possible $n$ for a given diameter and a given level of regularity. The scientific results in this field support our findings, too, because for $(k=3,d=2)$ the largest $n$ is 10, while for $(k=3,d=3)$ it is 20. In the case of $(k=4,d=2)$ the largest $n$ is 15, but for $(k=4,d=3)$ it is proved that the largest graph is much above our bound, but the optimal number of the vertices in this case is still an open question.

Finally, we can increase the regularity level to 5 in order to find out if we are able to get 2 as the smallest diameter for larger graphs. The answer is yes, actually it is also proven that $d=2$ is reachable for 5-regular graphs until 24 vertices, but of course we are interested in the specific graphs that could help us determine the adequate comparison patterns. Our results can be found in Table \hyperlink{tab:4}{4}. The $k=5$ parameter is only possible when $n$ is even again, so when it is odd, we let one vertex have 6 as its degree.

\begin{table}[ht!]
	\centering
	\renewcommand{\baselinestretch}{0.75}
	\small
	\begin{tabular}{|m{1cm}|m{3cm}|m{5cm}|}
	\hline \begin{center}	$\boldsymbol{k}$\textbf{=5} \end{center} & \begin{center} \hypertarget{tab:4}{Graph} \end{center}  & \begin{center} Further information \end{center} \\ \hline \hline

	\begin{center} $\boldsymbol{n}$\textbf{=16} \end{center} & \text{ }
	\raisebox{-\totalheight}{\includegraphics[width=0.15\textwidth,  height=0.05\paperheight]{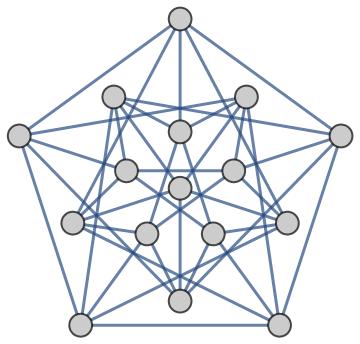}}  \begin{center} \footnotesize{Clebsch graph} \end{center}	 & \begin{itemize} \item 40/120 comparisons ($c\approx0.333$)  \item $\geq 3$ graphs \end{itemize} \\ \hline

    \begin{center} $\boldsymbol{n}$\textbf{=17} \end{center} & \text{ }
	\raisebox{-\totalheight}{\includegraphics[width=0.15\textwidth,  height=0.05\paperheight]{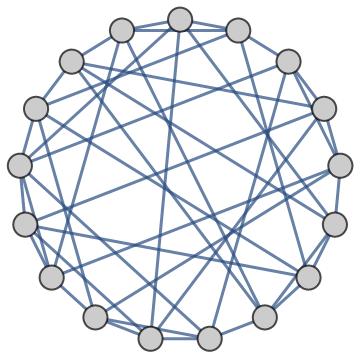}}  \begin{center} \footnotesize{} \end{center} & \begin{itemize} \item 43/136 comparisons  ($c\approx0.316$)\item $\geq 1$ graph \end{itemize} \\ \hline
	
    \begin{center} $\boldsymbol{n}$\textbf{=18} \end{center} & \text{ }
	\raisebox{-\totalheight}{\includegraphics[width=0.15\textwidth,  height=0.05\paperheight]{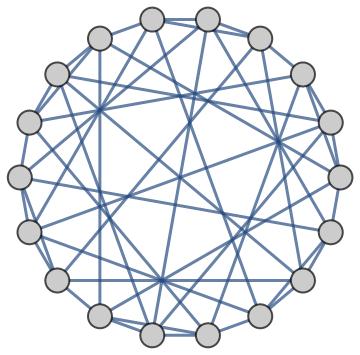}}  \begin{center} \footnotesize{(18,1)-noncayley transitive graph} \end{center}& \begin{itemize} \item 45/153 comparisons ($c\approx0.294$)  \item $\geq 1$ graph \end{itemize} \\ \hline
	
	\begin{center} $\boldsymbol{n}$\textbf{=19} \end{center} & \text{ }\raisebox{-\totalheight}{\includegraphics[width=0.15\textwidth,  height=0.05\paperheight]{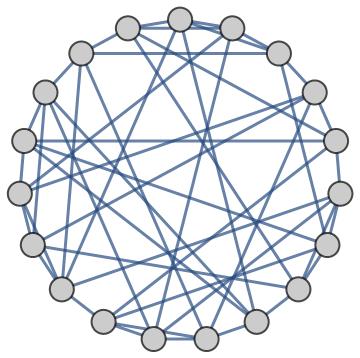}}  \begin{center} \footnotesize{} \end{center}
	 & \begin{itemize} \item 48/171 comparisons ($c\approx0.281$)  \item $\geq 1$ graph \end{itemize} \\ \hline

	 \begin{center}	$\boldsymbol{n}$\textbf{=20} \end{center} & \text{ } \raisebox{-\totalheight}{\includegraphics[width=0.15\textwidth,  height=0.05\paperheight]{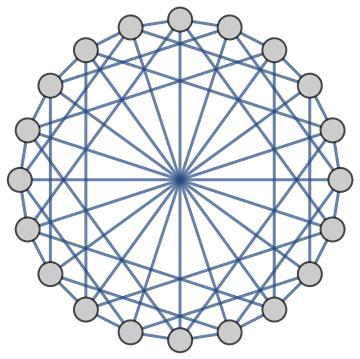}}  \begin{center} \footnotesize{(20,8)-noncayley transitive graph} \end{center} & \begin{itemize} \item 50/190 comparisons ($c\approx0.263$)  \item $\geq 1$ graph \end{itemize} \\ \hline
	\end{tabular}
	
	\renewcommand{\baselinestretch}{1}
	\Large
	\caption{$k=5$-(quasi-)regular graphs on $n$ vertices with minimal diameter $d=2$}
\end{table}

As we can see in this table there are higher completion ratios again, and for instance when we have 20 vertices the decision makers should make 50 comparisons which in certain situations can be too many. One can also note that in this table we report that there are some graphs with the needed properties, but never indicate the number of them. The reason behind this is simple: the really high number of the potential connected 5-regular graphs (for instance in the case of $n=20$ there are roughly $4\cdot10^{15}$ possibilities).

This means that we have examined all the cases that we previously called interesting. According to our results if we use the $(n,k,d)$ parameters, then for smaller MCDM problems the $k=3$ is enough to get 2 as the diameter of the representing graph which leads to a small completion ratio and a stable system of the comparisons. In larger problems, when we have more alternatives or criteria we can choose if we use $k=3$, when the completion ratio is smaller, but our approximation can be unstable, or choose higher regularity levels with more reliable results but a higher completion ratio. We also showed examples and graphs with the needed properties for the different cases, which can help anyone in a MCDM problem to decide which comparisons have to be made. One can find the summary of our results in Table ~\hyperlink{tab:5}{5}, which shows how many graphs we know for given $(n,k,d)$ parameters. It is also true that if there is a graph for $(n,k,d)$ in the table, then there are graphs for $(n,k,D)$ too, where $D>d$, and there is no graph with the parameters of $(n,k,d-1)$. We omitted the cases when $k=4$ and $n\leq10$, because the minimal diameter is the same as it was in the case of $k=3$. There is the same reasoning behind the emptiness of the table when $k=5$ and $n\leq15$. We have not included the cases when $k=4$ and $n\geq16$, because $d=3$ can be reached by $3$-regular graphs, but for $d=2$ at least $5$-regularity is \hypertarget{fig2}{needed}.

\begin{table}[h!]
\caption{The summary of the results: the number of $k$-(quasi-)regular graphs on $n$ nodes with diameter $d$}
	\begin{center} \includegraphics[width=0.4\textwidth,  height=0.4\paperheight]{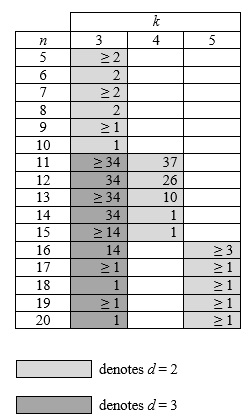}\\
 \end{center} 
\end{table}

\section{Conclusion and further research}

In this article we provided a systematic collection of recommended filling patterns of incomplete pairwise comparisons' using the graph representation of the PCMs. We discussed the applied methodology in many details, and then presented our results using the number of criteria or alternatives, the regularity level and the diameter of the representing graph as parameters. We showed that relatively small completion ratios can be achieved with small diameters, and provided examples for every case that we considered to be relevant.

The investigation of the robustness of the results, namely what is between the different regularity levels, could be the topic of a further research. It is also an interesting problem to concentrate directly on the completion ratio as a parameter instead of the regularity of the representing graph. If the $(n,c)$ pair is given then what comparisons are the most important to be made? We would like to deal with these questions in our future works.

Although our results have been presented within the framework of pairwise comparison matrices, they are applicable in a wider range. 
A lot of other models based on pairwise comparisons can utilize our findings. For example ranking of sport players or teams based on their matches
leads to the problem of tournament design: which pairs should play against each other?

\bibliographystyle{apalike}
\bibliography{bszt}
\addcontentsline{toc}{section}{References}

\setlength{\headheight}{14.5pt}

%\begin{document}

\section{Appendix1: $k=3$ and $d=2$ graphs}

\begin{figure}[h]
\caption{$n=5$, $k=3$, $d=2$ example}
\begin{center} \includegraphics[width=0.4\textwidth,  height=0.2\paperheight]{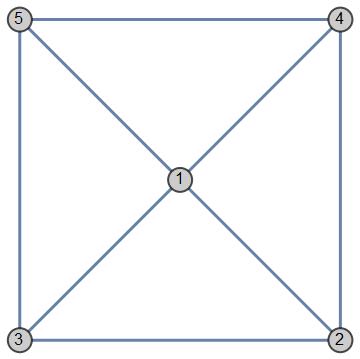}\\
'Graph6' format: \copyablespace{D\}k}\end{center} 
\end{figure}

\begin{table}[ht!]
\centering
\renewcommand{\baselinestretch}{1.25} 
\normalsize
\begin{tabular}{|m{0.3cm}|m{0.3cm}|m{0.3cm}|m{0.3cm}|m{0.3cm}|m{0.3cm}|} 

\hline
\text{ }& {1} & {2} & {3} & {4} & {5} \\ \hline
{1} & \cellcolor{gray}  & $\bullet$ & $\bullet$ & $\bullet$ & $\bullet$ \\ \hline
{2} & $\circ$ & \cellcolor{gray} & $\bullet$ & $\bullet$ &  \\ \hline
{3} & $\circ$ & $\circ$ & \cellcolor{gray} &  & $\bullet$  \\ \hline
{4} & $\circ$ & $\circ$ & & \cellcolor{gray} & $\bullet$  \\ \hline
{5} & $\circ$ &  & $\circ$ & $\circ$  & \cellcolor{gray} \\ \hline

\end{tabular}
	\renewcommand{\baselinestretch}{1} 
	\normalsize
	\caption{The bullets in the matrix shows the edges of the graph}
	\end{table}

	Edges
\begin{multicols*}{2}	
	1-2
	
	1-3
	
	1-4
 
 1-5

 2-3
 
 2-4
 
 3-5
  
 4-5

\end{multicols*}

\newpage

\begin{figure}[h]
\caption{$n=6$, $k=3$, $d=2$ example: 3-prism graph}
\begin{center} \includegraphics[width=0.4\textwidth,  height=0.2\paperheight]{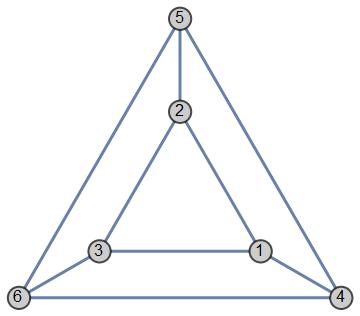}\\
'Graph6' format: \copyablespace{E\{Sw} \end{center} 
\end{figure}

\begin{table}[ht!]
\centering
\renewcommand{\baselinestretch}{1.25} 
\normalsize
\begin{tabular}{|m{0.3cm}|m{0.3cm}|m{0.3cm}|m{0.3cm}|m{0.3cm}|m{0.3cm}|m{0.3cm}|} 

\hline
\text{ }& {1} & {2} & {3} & {4} & {5} & {6} \\ \hline
{1} & \cellcolor{gray}  & $\bullet$ & $\bullet$ & $\bullet$ & & \\ \hline
{2} & $\circ$ & \cellcolor{gray} & $\bullet$ & & $\bullet$ &   \\ \hline
{3} & $\circ$ & $\circ$ & \cellcolor{gray} &  & &  $\bullet$ \\ \hline
{4} & $\circ$ & & & \cellcolor{gray} & $\bullet$ &  $\bullet$  \\ \hline
{5} & &  $\circ$  &  & $\circ$  & \cellcolor{gray} &  $\bullet$ \\ \hline
{6} &  &  & $\circ$ & $\circ$  &  $\circ$ & \cellcolor{gray}  \\ \hline
\end{tabular}
	\renewcommand{\baselinestretch}{1} 
	\normalsize
	\caption{The bullets in the matrix shows the edges of the graph}
	\end{table}

	Edges
\begin{multicols*}{2}	
	1-2
	
	1-3
	
	1-4
 
 2-3

 2-5
 
 3-6
 
 4-5
  
 4-6
 
 5-6

\end{multicols*}

\newpage

\begin{figure}[h]
\caption{$n=7$, $k=3$, $d=2$ example}
\begin{center} \includegraphics[width=0.4\textwidth,  height=0.2\paperheight]{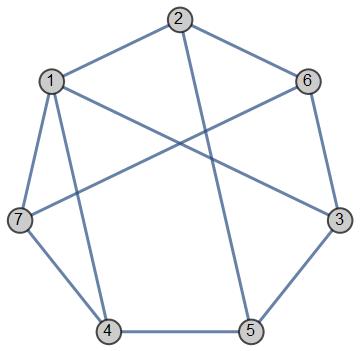}\\
'Graph6' format: \copyablespace{FsdrO} \end{center} 
\end{figure}

\begin{table}[ht!]
\centering
\renewcommand{\baselinestretch}{1.25} 
\normalsize
\begin{tabular}{|m{0.3cm}|m{0.3cm}|m{0.3cm}|m{0.3cm}|m{0.3cm}|m{0.3cm}|m{0.3cm}|m{0.3cm}|} 

\hline
\text{ }& {1} & {2} & {3} & {4} & {5} & {6} & {7} \\ \hline
{1} & \cellcolor{gray}  & $\bullet$ & $\bullet$ & $\bullet$ & & & $\bullet$ \\ \hline
{2} & $\circ$ & \cellcolor{gray} & & & $\bullet$ & $\bullet$ &  \\ \hline
{3} & $\circ$ & & \cellcolor{gray} &  & $\bullet$ &  $\bullet$ & \\ \hline
{4} & $\circ$ & & & \cellcolor{gray} & $\bullet$ &  & $\bullet$ \\ \hline
{5} & &  $\circ$  & $\circ$  & $\circ$  & \cellcolor{gray} & & \\ \hline
{6} &  & $\circ$ & $\circ$ & & & \cellcolor{gray} & $\bullet$ \\ \hline
{7} & $\circ$  &  & & $\circ$  & & $\circ$ & \cellcolor{gray} \\ \hline
\end{tabular}
	\renewcommand{\baselinestretch}{1} 
	\normalsize
	\caption{The bullets in the matrix shows the edges of the graph}
	\end{table}

	Edges
\begin{multicols*}{2}	
	1-2
	
	1-3
	
	1-4
	
	1-7
 
 2-5

 2-6
 
 3-5
 
 3-6
 
 4-5
  
 4-7
 
 6-7
 
\end{multicols*}

\newpage

\begin{figure}[h]
\caption{$n=8$, $k=3$, $d=2$ example: Wagner graph}
\begin{center} \includegraphics[width=0.4\textwidth,  height=0.2\paperheight]{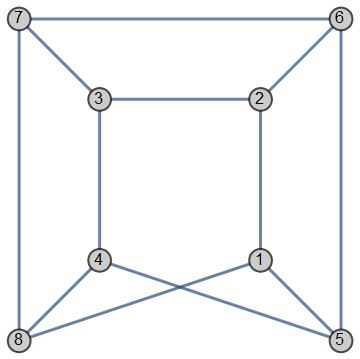}\\
'Graph6' format: \copyablespace{GhdHKc}  \end{center} 
\end{figure}

\begin{table}[ht!]
\centering
\renewcommand{\baselinestretch}{1.25} 
\normalsize
\begin{tabular}{|m{0.3cm}|m{0.3cm}|m{0.3cm}|m{0.3cm}|m{0.3cm}|m{0.3cm}|m{0.3cm}|m{0.3cm}|m{0.3cm}|} 

\hline
\text{ }& {1} & {2} & {3} & {4} & {5} & {6} & {7} & {8} \\ \hline
{1} & \cellcolor{gray}  & $\bullet$ & & & $\bullet$ & & & $\bullet$ \\ \hline
{2} & $\circ$ & \cellcolor{gray} & $\bullet$ & & & $\bullet$ & & \\ \hline
{3} & & $\circ$ & \cellcolor{gray} & $\bullet$ & & & $\bullet$ & \\ \hline
{4} & & & $\circ$ & \cellcolor{gray} & $\bullet$ &  & & $\bullet$ \\ \hline
{5} & $\circ$ &  & & $\circ$  & \cellcolor{gray} & $\bullet$ & & \\ \hline
{6} &  & $\circ$ & & & $\circ$ & \cellcolor{gray} & $\bullet$ & \\ \hline
{7} & &  & $\circ$ &  & & $\circ$ & \cellcolor{gray} & $\bullet$ \\ \hline
{8} & $\circ$  &  & & $\circ$  & & & $\circ$ &  \cellcolor{gray} \\ \hline
\end{tabular}
	\renewcommand{\baselinestretch}{1} 
	\normalsize
	\caption{The bullets in the matrix shows the edges of the graph}
	\end{table}

	Edges
\begin{multicols*}{3}	
	1-2
	
	1-5
	
	1-8
	
	2-3
 
 2-6

 3-4
 
 3-7
 
 4-5
 
 4-8
  
 5-6
 
 6-7
 
 7-8
 
\end{multicols*}

\newpage

\begin{figure}[h]
\caption{$n=9$, $k=3$, $d=2$ example}
\begin{center} \includegraphics[width=0.4\textwidth,  height=0.2\paperheight]{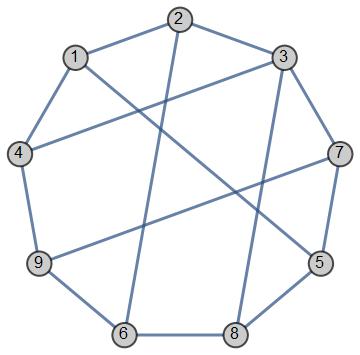}\\
'Graph6' format: \copyablespace{HsT@PWU} \end{center} 
\end{figure}

\begin{table}[ht!]
\centering
\renewcommand{\baselinestretch}{1.25} 
\normalsize
\begin{tabular}{|m{0.3cm}|m{0.3cm}|m{0.3cm}|m{0.3cm}|m{0.3cm}|m{0.3cm}|m{0.3cm}|m{0.3cm}|m{0.3cm}|m{0.3cm}|} 

\hline
\text{ }& {1} & {2} & {3} & {4} & {5} & {6} & {7} & {8} & {9} \\ \hline
{1} & \cellcolor{gray}  & $\bullet$ & & $\bullet$ & $\bullet$ & & & & \\ \hline
{2} & $\circ$ & \cellcolor{gray} & $\bullet$ & & & $\bullet$ & & & \\ \hline
{3} & & $\circ$ & \cellcolor{gray} & $\bullet$ & & & $\bullet$ & $\bullet$ & \\ \hline
{4} & $\circ$ & & $\circ$ & \cellcolor{gray} & &  & & & $\bullet$ \\ \hline
{5} & $\circ$ &  & &  & \cellcolor{gray} & & $\bullet$ & $\bullet$ & \\ \hline
{6} &  & $\circ$ & & & & \cellcolor{gray} & & $\bullet$ & $\bullet$ \\ \hline
{7} & &  & $\circ$ &  & $\circ$ & & \cellcolor{gray} & & $\bullet$ \\ \hline
{8} &  &  & $\circ$ & & $\circ$ & $\circ$ & &  \cellcolor{gray} & \\ \hline
{9} &  &  & & $\circ$  & & $\circ$ & $\circ$ & & \cellcolor{gray} \\ \hline
\end{tabular}
	\renewcommand{\baselinestretch}{1} 
	\normalsize
	\caption{The bullets in the matrix shows the edges of the graph}
	\end{table}

\newpage
	Edges
\begin{multicols*}{2}	
	1-2
	
	1-4
	
	1-5
	
	2-3
 
 2-6

 3-4
 
 3-7
 
  3-8
 
 4-9
 
 5-7
  
 5-8
 
 6-8
 
 6-9
 
 7-9
 
\end{multicols*}

\newpage

\begin{figure}[h]
\caption{$n=10$, $k=3$, $d=2$ example: Petersen graph}
\begin{center} \includegraphics[width=0.4\textwidth,  height=0.2\paperheight]{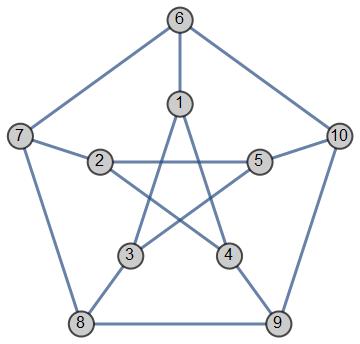}\\
'Graph6' format: \copyablespace{IUYAHCPBG} \end{center} 
\end{figure}

\begin{table}[ht!]
\centering
\renewcommand{\baselinestretch}{1.25} 
\normalsize
\begin{tabular}{|m{0.4cm}|m{0.3cm}|m{0.3cm}|m{0.3cm}|m{0.3cm}|m{0.3cm}|m{0.3cm}|m{0.3cm}|m{0.3cm}|m{0.3cm}|m{0.3cm}|} 

\hline
\text{ }& {1} & {2} & {3} & {4} & {5} & {6} & {7} & {8} & {9} & {10}\\ \hline
{1} & \cellcolor{gray}  & & $\bullet$ & $\bullet$ & & $\bullet$ & & & & \\ \hline
{2} & & \cellcolor{gray} & & $\bullet$ & $\bullet$ &  & $\bullet$ & & &\\ \hline
{3} & $\circ$ & & \cellcolor{gray} &  & $\bullet$ &  & & $\bullet$ & & \\ \hline
{4} & $\circ$ & $\circ$ & & \cellcolor{gray} & & & & & $\bullet$ & \\ \hline
{5} & & $\circ$ & $\circ$ &  & \cellcolor{gray} &  & & & & $\bullet$ \\ \hline
{6} & $\circ$ & & &  & & \cellcolor{gray} & $\bullet$ & & & $\bullet$ \\ \hline
{7} & & $\circ$ &  & &  & $\circ$ & \cellcolor{gray} & $\bullet$ & & \\ \hline
{8} & &  & $\circ$ &  & & & $\circ$ & \cellcolor{gray}  & $\bullet$ & \\ \hline
{9} & & & & $\circ$ & & & & $\circ$ & \cellcolor{gray} & $\bullet$ \\ \hline
{10} & &  & & & $\circ$  & $\circ$ & &  & $\circ$ & \cellcolor{gray}\\ \hline

\end{tabular}
	\renewcommand{\baselinestretch}{1} 
	\normalsize
	\caption{The bullets in the matrix shows the edges of the graph}
	\end{table}
	
	\newpage
	Edges
\begin{multicols*}{2}	
	1-3
	
	1-4
	
	1-6
 
 2-4

 2-5
 
 2-7
 
 3-5
  
 3-8
 
 4-9
 
 5-10
 
 6-7
 
 6-10
  
 7-8
 
 8-9
 
 9-10
 
\end{multicols*}

\newpage

\section{Appendix2: $k=3$ and $d=3$ graphs}
\begin{figure}[h]
\caption{$n=11$, $k=3$, $d=3$ example}
\begin{center} \includegraphics[width=0.4\textwidth,  height=0.2\paperheight]{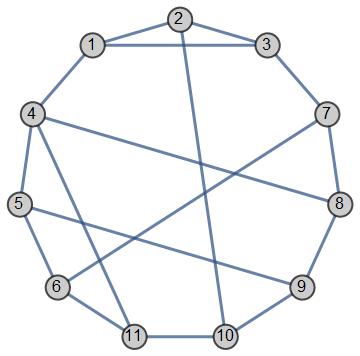}\\
'Graph6' format: \copyablespace{J\{COXCPAIG\_} \end{center} 
\end{figure}

\begin{table}[ht!]
\centering
\renewcommand{\baselinestretch}{1.25} 
\normalsize
\begin{tabular}{|m{0.4cm}|m{0.3cm}|m{0.3cm}|m{0.3cm}|m{0.3cm}|m{0.3cm}|m{0.3cm}|m{0.3cm}|m{0.3cm}|m{0.3cm}|m{0.3cm}|m{0.3cm}|} 

\hline
\text{ }& {1} & {2} & {3} & {4} & {5} & {6} & {7} & {8} & {9} & {10} & {11}\\ \hline
{1} & \cellcolor{gray}  & $\bullet$ & $\bullet$ & $\bullet$ & & & & & & & \\ \hline
{2} & $\circ$ & \cellcolor{gray} & $\bullet$ & & &  & & & & $\bullet$ &\\ \hline
{3} & $\circ$ & $\circ$ & \cellcolor{gray} &  & &  & $\bullet$ & & & & \\ \hline
{4} & $\circ$ & & & \cellcolor{gray} & $\bullet$ & & & $\bullet$ & & & $\bullet$ \\ \hline
{5} & & & & $\circ$  & \cellcolor{gray} & $\bullet$ & & & $\bullet$ & & \\ \hline
{6} & & & &  & $\circ$ & \cellcolor{gray} & $\bullet$ & & & & $\bullet$ \\ \hline
{7} & & & $\circ$  & &  & $\circ$ & \cellcolor{gray} & $\bullet$ & & & \\ \hline
{8} & &  &  & $\circ$ & & & $\circ$ & \cellcolor{gray}  & $\bullet$ & & \\ \hline
{9} & & & & & $\circ$  & & & $\circ$ & \cellcolor{gray} & $\bullet$ & \\ \hline
{10} & & $\circ$ & & &  &  & &  & $\circ$ & \cellcolor{gray} & $\bullet$ \\ \hline
{11} & &  & & $\circ$  & & $\circ$ & &  & & $\circ$ & \cellcolor{gray}\\ \hline

\end{tabular}
	\renewcommand{\baselinestretch}{1} 
	\normalsize
	\caption{The bullets in the matrix shows the edges of the graph}
	\end{table}
	
	\newpage
	Edges
\begin{multicols*}{2}	
	1-2
	
	1-3
	
	1-4
 
 2-3

 2-10
 
 3-7
 
 4-5
  
 4-8
 
 4-11
 
 5-6
 
 5-9
 
 6-7
 
 6-11
  
 7-8
 
 8-9
 
 9-10
 
 10-11
 
\end{multicols*}

\newpage
\begin{figure}[h]
\caption{$n=12$, $k=3$, $d=3$ example: Tietze graph}
\begin{center} \includegraphics[width=0.4\textwidth,  height=0.2\paperheight]{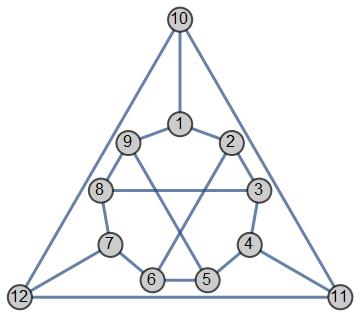}\\
'Graph6' format: \copyablespace{KhDGHEH\_?\_\_R} \end{center} 
\end{figure}

\begin{table}[ht!]
\centering
\renewcommand{\baselinestretch}{1.25} 
\normalsize
\begin{tabular}{|m{0.4cm}|m{0.3cm}|m{0.3cm}|m{0.3cm}|m{0.3cm}|m{0.3cm}|m{0.3cm}|m{0.3cm}|m{0.3cm}|m{0.3cm}|m{0.3cm}|m{0.3cm}|m{0.3cm}|} 

\hline
\text{ }& {1} & {2} & {3} & {4} & {5} & {6} & {7} & {8} & {9} & {10}  & {11} & {12}\\ \hline
{1} & \cellcolor{gray}  & $\bullet$ & &  & & &  & & $\bullet$ & $\bullet$ & & \\ \hline
{2} & $\circ$ & \cellcolor{gray} & $\bullet$ & &  & $\bullet$ & & & & & &\\ \hline
{3} & & $\circ$ & \cellcolor{gray} & $\bullet$ & & & & $\bullet$ & & & & \\ \hline
{4} & & & $\circ$ & \cellcolor{gray} & $\bullet$ & & & & & & $\bullet$ & \\ \hline
{5} & & & & $\circ$ & \cellcolor{gray} & $\bullet$ & & & $\bullet$ & & & \\ \hline
{6} & & $\circ$ & & & $\circ$ & \cellcolor{gray} & $\bullet$ & & & & & \\ \hline
{7} & & & & & & $\circ$ & \cellcolor{gray} & $\bullet$ & & & & $\bullet$ \\ \hline
{8} & &  & $\circ$ & & & & $\circ$ & \cellcolor{gray} & $\bullet$ & & & \\ \hline
{9} & $\circ$ & & & & $\circ$ & & & $\circ$ & \cellcolor{gray} & & & \\ \hline
{10} & $\circ$ & & & & &  & & & & \cellcolor{gray} & $\bullet$ & $\bullet$\\ \hline
{11} & & & & $\circ$ & & & & & & $\circ$ & \cellcolor{gray} & $\bullet$\\ \hline
{12} & & & & & & & $\circ$ &  & & $\circ$ & $\circ$ & \cellcolor{gray}\\ \hline

\end{tabular}
	\renewcommand{\baselinestretch}{1} 
	\normalsize
	\caption{The bullets in the matrix shows the edges of the graph}
	\end{table}
	
	\newpage
	
	Edges

\begin{multicols*}{2}

	1-2
	
	1-9
	
	1-10
 
 2-3

 2-6
 
 3-4
  
 3-8
 
 4-5

 4-11
 
 5-6
 
 5-9
 
 6-7
 
 7-8
  
 7-12
 
 8-9
 
 10-11
 
 10-12
 
 11-12
 
 \end{multicols*}

\newpage

\begin{figure}[h]
\caption{$n=13$, $k=3$, $d=3$ example}
\begin{center} \includegraphics[width=0.4\textwidth,  height=0.2\paperheight]{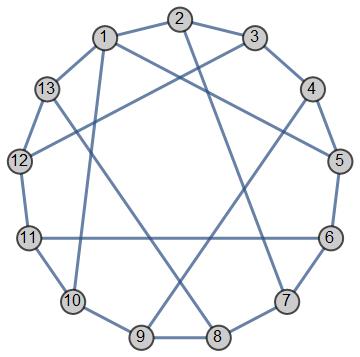}\\
'Graph6' format: \copyablespace{LhcIGCP\_GGc@\_P} \end{center} 
\end{figure}

\begin{table}[ht!]
\centering
\renewcommand{\baselinestretch}{1.25} 
\normalsize
\begin{tabular}{|m{0.4cm}|m{0.3cm}|m{0.3cm}|m{0.3cm}|m{0.3cm}|m{0.3cm}|m{0.3cm}|m{0.3cm}|m{0.3cm}|m{0.3cm}|m{0.3cm}|m{0.3cm}|m{0.3cm}|m{0.3cm}|} 

\hline
\text{ }& {1} & {2} & {3} & {4} & {5} & {6} & {7} & {8} & {9} & {10}  & {11} & {12} & {13}\\ \hline
{1} & \cellcolor{gray}  & $\bullet$ & &  & $\bullet$ & &  & & & $\bullet$ & & & $\bullet$ \\ \hline
{2} & $\circ$ & \cellcolor{gray} & $\bullet$ & &  &  & $\bullet$ & & & & & &\\ \hline
{3} & & $\circ$ & \cellcolor{gray} & $\bullet$ & & & & & & & & $\bullet$ & \\ \hline
{4} & & & $\circ$ & \cellcolor{gray} & $\bullet$ & & & & $\bullet$ & & & & \\ \hline
{5} & $\circ$ & & & $\circ$ & \cellcolor{gray} & $\bullet$ & & & & & & & \\ \hline
{6} & & & & & $\circ$ & \cellcolor{gray} & $\bullet$ & & & & $\bullet$ & & \\ \hline
{7} & & $\circ$ & & & & $\circ$ & \cellcolor{gray} & $\bullet$ & & & & & \\ \hline
{8} & &  & & & & & $\circ$ & \cellcolor{gray} & $\bullet$ & & & & $\bullet$ \\ \hline
{9} & & & & $\circ$ & & & & $\circ$ & \cellcolor{gray} & $\bullet$ & & & \\ \hline
{10} & $\circ$ & & & & &  & & & $\circ$ & \cellcolor{gray} & $\bullet$ & & \\ \hline
{11} & & & & & & $\circ$ & & & & $\circ$ & \cellcolor{gray} & $\bullet$ &\\ \hline
{12} & & & $\circ$ & & & & &  & & & $\circ$ & \cellcolor{gray} & $\bullet$ \\ \hline
{13} & $\circ$ & & & & & & & $\circ$  & & & & $\circ$ & \cellcolor{gray} \\ \hline

\end{tabular}
	\renewcommand{\baselinestretch}{1} 
	\normalsize
	\caption{The bullets in the matrix shows the edges of the graph}
	\end{table}
	
	\newpage
	
	Edges

\begin{multicols*}{2}

	1-2
	
	1-5
	
	1-10
	
	1-13
 
 2-3

 2-7
 
 3-4
  
 3-12
 
 4-5

 4-9
 
 5-6
 
 6-7
 
 6-11
 
 7-8
  
 8-9
 
 8-13
 
 9-10
 
 10-11
 
 11-12
 
 12-13
 
 \end{multicols*}

\newpage

\begin{figure}[h]
\caption{$n=14$, $k=3$, $d=3$ example: Heawood graph}
\begin{center} \includegraphics[width=0.4\textwidth,  height=0.2\paperheight]{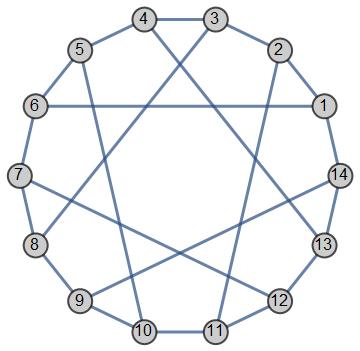}\\
'Graph6' format: \copyablespace{MhEGHC@AI?\_PC@\_G\_} \end{center} 
\end{figure}

\begin{table}[ht!]
\centering
\renewcommand{\baselinestretch}{1.25} 
\normalsize
\begin{tabular}{|m{0.4cm}|m{0.3cm}|m{0.3cm}|m{0.3cm}|m{0.3cm}|m{0.3cm}|m{0.3cm}|m{0.3cm}|m{0.3cm}|m{0.3cm}|m{0.3cm}|m{0.3cm}|m{0.3cm}|m{0.3cm}|m{0.3cm}|} 

\hline
\text{ }& {1} & {2} & {3} & {4} & {5} & {6} & {7} & {8} & {9} & {10}  & {11} & {12} & {13} & {14}\\ \hline
{1} & \cellcolor{gray}  & $\bullet$ & &  & & $\bullet$ &  & & &  & & & & $\bullet$ \\ \hline
{2} & $\circ$ & \cellcolor{gray} & $\bullet$ & &  &  & & & & & $\bullet$ & & &\\ \hline
{3} & & $\circ$ & \cellcolor{gray} & $\bullet$ & & & & $\bullet$ & & & & & & \\ \hline
{4} & & & $\circ$ & \cellcolor{gray} & $\bullet$ & & & & & & & & $\bullet$ & \\ \hline
{5} & & & & $\circ$ & \cellcolor{gray} & $\bullet$ & & & & $\bullet$ & & & & \\ \hline
{6} & $\circ$ & & & & $\circ$ & \cellcolor{gray} & $\bullet$ & & & & & & & \\ \hline
{7} & &  & & & & $\circ$ & \cellcolor{gray} & $\bullet$ & & & & $\bullet$ & & \\ \hline
{8} & &  & $\circ$ & & & & $\circ$ & \cellcolor{gray} & $\bullet$ & & & & & \\ \hline
{9} & & & & & & & & $\circ$ & \cellcolor{gray} & $\bullet$ & & & & $\bullet$ \\ \hline
{10} & & & & & $\circ$ &  & & & $\circ$ & \cellcolor{gray} & $\bullet$ & & & \\ \hline
{11} & & $\circ$ & & & & & & & & $\circ$ & \cellcolor{gray} & $\bullet$ & &\\ \hline
{12} & & & & & & & $\circ$ &  & & & $\circ$ & \cellcolor{gray} & $\bullet$ & \\ \hline
{13} & & & & $\circ$ & & & &  & & & & $\circ$ & \cellcolor{gray} & $\bullet$ \\ \hline
{14} & $\circ$ & & & & & & & & $\circ$  & & & & $\circ$  & \cellcolor{gray} \\ \hline

\end{tabular}
	\renewcommand{\baselinestretch}{1} 
	\normalsize
	\caption{The bullets in the matrix shows the edges of the graph}
	\end{table}
	
	\newpage
	
	Edges

\begin{multicols*}{2}

	1-2
	
	1-6
	
	1-14
	
	2-3
 
 2-11

 3-4
 
 3-8
  
 4-5
 
 4-13

 5-6
 
 5-10
 
 6-7
 
 7-8
 
 7-12
  
 8-9
 
 9-10
 
 9-14
 
 10-11
 
 11-12
 
 12-13
 
 13-14
 
 \end{multicols*}

\newpage

\begin{figure}[h]
\caption{$n=15$, $k=3$, $d=3$ example}
\begin{center} \includegraphics[width=0.4\textwidth,  height=0.19\paperheight]{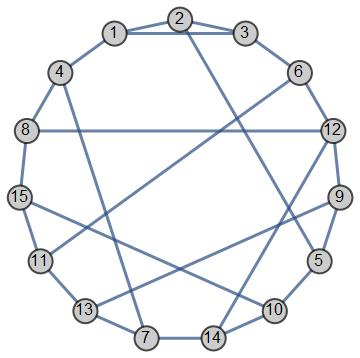}\\
'Graph6' format: \copyablespace{N\{O\_\_\_GA?G?k?i?d?J?} \end{center} 
\end{figure}

\begin{table}[ht!]
\centering
\renewcommand{\baselinestretch}{1.25} 
\normalsize
\begin{tabular}{|m{0.4cm}|m{0.3cm}|m{0.3cm}|m{0.3cm}|m{0.3cm}|m{0.3cm}|m{0.3cm}|m{0.3cm}|m{0.3cm}|m{0.3cm}|m{0.3cm}|m{0.3cm}|m{0.3cm}|m{0.3cm}|m{0.3cm}|m{0.3cm}|} 

\hline
\text{ }& {1} & {2} & {3} & {4} & {5} & {6} & {7} & {8} & {9} & {10}  & {11} & {12} & {13} & {14} & {15}\\ \hline
{1} & \cellcolor{gray}  & $\bullet$ & $\bullet$ & $\bullet$  & &  &  & & &  & & & & & \\ \hline
{2} & $\circ$ & \cellcolor{gray} & $\bullet$ & & $\bullet$ &  & & & & & & & & &\\ \hline
{3} & $\circ$ & $\circ$ & \cellcolor{gray} &  & &$\bullet$ & & & & & & & & &\\ \hline
{4} & $\circ$ & & & \cellcolor{gray} &  & & $\bullet$ & $\bullet$ & & & & & & &\\ \hline
{5} & & $\circ$ & & & \cellcolor{gray} & & & & $\bullet$ & $\bullet$ & & & & &\\ \hline
{6} & & & $\circ$ & & & \cellcolor{gray} & & & & & $\bullet$ & $\bullet$ &  & & \\ \hline
{7} & &  & & $\circ$ & & & \cellcolor{gray} & & & & & & $\bullet$ & $\bullet$ &\\ \hline
{8} & &  &  & $\circ$ & & & & \cellcolor{gray} & & & & $\bullet$ & & & $\bullet$ \\ \hline
{9} & & & & & $\circ$ & & & & \cellcolor{gray} & & & $\bullet$ & $\bullet$ & & \\ \hline
{10} & & & & & $\circ$ &  & & & & \cellcolor{gray} & & & & $\bullet$ & $\bullet$ \\ \hline
{11} & & & & & & $\circ$ & & & & & \cellcolor{gray} & & $\bullet$ & & $\bullet$ \\ \hline
{12} & & & & & & $\circ$ & & $\circ$ & $\circ$ & & & \cellcolor{gray} & & $\bullet$ &  \\ \hline
{13} & & & & & & & $\circ$ &  & $\circ$ & & $\circ$ & & \cellcolor{gray} & & \\ \hline
{14} & & & & & & & $\circ$ & &  & $\circ$ &  & $\circ$ & & \cellcolor{gray} & \\ \hline
{15} & & & & & & & & $\circ$ &  & $\circ$ & $\circ$ & & & & \cellcolor{gray} \\ \hline

\end{tabular}
	\renewcommand{\baselinestretch}{1} 
	\normalsize
	\caption{The bullets in the matrix shows the edges of the graph}
	\end{table}
	
	\newpage
	
	Edges

\begin{multicols*}{2}

	1-2
	
	1-3
	
	1-4
	
	2-3
 
 2-5
 
 3-6
 
 4-7
 
 4-8
  
 5-9
 
 5-10
 
 6-11
 
 6-12
 
 7-13
 
 7-14
 
 8-12
  
 8-15
 
 9-12
 
 9-13
 
 10-14
 
 10-15
 
 11-13
 
 11-15
 
 12-14
 
 \end{multicols*}
 
 \newpage

\begin{figure}[h]
\caption{$n=16$, $k=3$, $d=3$ example}
\begin{center} \includegraphics[width=0.4\textwidth,  height=0.19\paperheight]{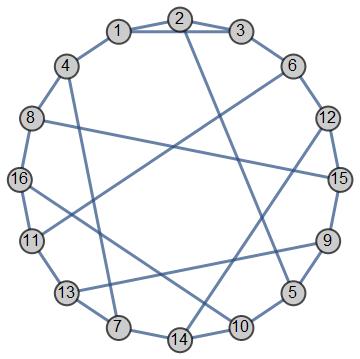}\\
'Graph6' format: \copyablespace{O\{O\_\_\_GA?G?\_?i?d?K\_Ao} \end{center} 
\end{figure}

\begin{table}[ht!]
\centering
\renewcommand{\baselinestretch}{1.15} 
\normalsize
\begin{tabular}{|m{0.4cm}|m{0.276cm}|m{0.276cm}|m{0.276cm}|m{0.276cm}|m{0.276cm}|m{0.276cm}|m{0.276cm}|m{0.276cm}|m{0.276cm}|m{0.276cm}|m{0.276cm}|m{0.276cm}|m{0.276cm}|m{0.276cm}|m{0.276cm}|m{0.276cm}|} 

\hline
\text{ }& {1} & {2} & {3} & {4} & {5} & {6} & {7} & {8} & {9} & {10}  & {11} & {12} & {13} & {14} & {15}& {16}\\ \hline
{1} & \cellcolor{gray}  & $\bullet$ &  $\bullet$ & $\bullet$ & & &  & & &  & & & & & & \\ \hline
{2} & $\circ$ & \cellcolor{gray} & $\bullet$ & & $\bullet$ &  & & & & & & & & & & \\ \hline
{3} & $\circ$ & $\circ$  & \cellcolor{gray} &  & & $\bullet$ & & & & & & & & & &\\ \hline
{4} & $\circ$ & & & \cellcolor{gray} & & & $\bullet$ & $\bullet$ & & & & & & & &\\ \hline
{5} & & $\circ$ & & & \cellcolor{gray} & & & & $\bullet$ & $\bullet$ & &  & & & &\\ \hline
{6} & & & $\circ$ & & & \cellcolor{gray} & & & & &$\bullet$ & $\bullet$ & & & & \\ \hline
{7} &  &  & & $\circ$ & & & \cellcolor{gray} & & & & & & $\bullet$ & $\bullet$ & & \\ \hline
{8} & & & & $\circ$ & & & & \cellcolor{gray} & & & & & & & $\bullet$ & $\bullet$ \\ \hline
{9} & & & & & $\circ$ & & & & \cellcolor{gray} & & & & $\bullet$ & & $\bullet$ & \\ \hline
{10} & & & & & $\circ$ &  & & & & \cellcolor{gray} & & & & $\bullet$ & & $\bullet$ \\ \hline
{11} & & & & & & $\circ$ & & & & & \cellcolor{gray} & & $\bullet$ & & & $\bullet$ \\ \hline
{12} & & & & & & $\circ$ & & & & & & \cellcolor{gray} & & $\bullet$ & $\bullet$ & \\ \hline
{13} & & & & & & & $\circ$ &  & $\circ$ & & $\circ$ & & \cellcolor{gray} & & & \\ \hline
{14} & & & & & & & $\circ$ & &  & $\circ$ &  & $\circ$ & & \cellcolor{gray} & & \\ \hline
{15} & & & & & & & & $\circ$ & $\circ$ & & & $\circ$ &  & & \cellcolor{gray} & \\ \hline
{16} & & & & & & & & $\circ$ &  & $\circ$ & $\circ$ & &  & & & \cellcolor{gray} \\ \hline

\end{tabular}
	\renewcommand{\baselinestretch}{1} 
	\normalsize
	\caption{The bullets in the matrix shows the edges of the graph}
	\end{table}
	
	\newpage
	
	Edges

\begin{multicols*}{2}

	1-2
	
	1-3
	
	1-4
	
	2-3
 
 2-5
 
 3-6
 
 4-7
 
 4-8
  
 5-9
 
 5-10
 
 6-11
 
 6-12
 
 7-13
 
 7-14
 
 8-15
  
 8-16
 
 9-13
 
 9-15
 
 10-14
 
 10-16
 
 11-13
 
 11-16
 
 12-14
 
 12-15
 
 \end{multicols*}
 
 \newpage
\begin{figure}[h]
\caption{$n=17$, $k=3$, $d=3$ example}
\begin{center} \includegraphics[width=0.4\textwidth,  height=0.19\paperheight]{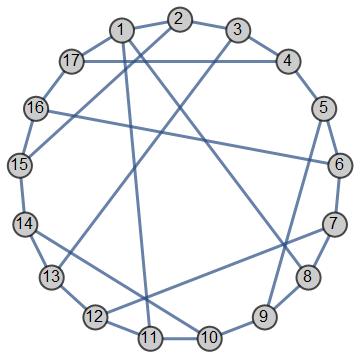}\\
'Graph6' format: \copyablespace{PhCGKCH?K?\_PG@?Cg?GG@c?C} \end{center} 
\end{figure}

\begin{table}[ht!]
\centering
\renewcommand{\baselinestretch}{1.05} 
\normalsize
\begin{tabular}{|m{0.3cm}|m{0.252cm}|m{0.252cm}|m{0.252cm}|m{0.252cm}|m{0.252cm}|m{0.252cm}|m{0.252cm}|m{0.252cm}|m{0.252cm}|m{0.252cm}|m{0.252cm}|m{0.252cm}|m{0.252cm}|m{0.252cm}|m{0.252cm}|m{0.252cm}|m{0.252cm}|} 

\hline
\text{ }& {1} & {2} & {3} & {4} & {5} & {6} & {7} & {8} & {9} & {10}  & {11} & {12} & {13} & {14} & {15}& {16} & {17}\\ \hline
{1} & \cellcolor{gray}  & $\bullet$ &  &  & & &  & $\bullet$ & &  & $\bullet$ & & & & & & $\bullet$ \\ \hline
{2} & $\circ$ & \cellcolor{gray} & $\bullet$ & &  &  & & & & & & & & & $\bullet$ & & \\ \hline
{3} & & $\circ$  & \cellcolor{gray} & $\bullet$ & & & & & & & & & $\bullet$ & & & & \\ \hline
{4} & & & $\circ$ & \cellcolor{gray} & $\bullet$ & & & & & & & & & & & & $\bullet$ \\ \hline
{5} & & & & $\circ$ & \cellcolor{gray} & $\bullet$ & & & $\bullet$ & & &  & & & & & \\ \hline
{6} & & & & & $\circ$ & \cellcolor{gray} & $\bullet$ & & & & &  & & & & $\bullet$ &  \\ \hline
{7} &  &  & & & & $\circ$ & \cellcolor{gray} & $\bullet$ & & & & $\bullet$ & & & & & \\ \hline
{8} & $\circ$ & & & & & & $\circ$ & \cellcolor{gray} & $\bullet$ & & & & & & & & \\ \hline
{9} & & & & & $\circ$ & & & $\circ$ & \cellcolor{gray} & $\bullet$ & & & & & & & \\ \hline
{10} & & & & & &  & & & $\circ$ & \cellcolor{gray} & $\bullet$ & & & $\bullet$ & & & \\ \hline
{11} & $\circ$ & & & & & & & & & $\circ$ & \cellcolor{gray} & $\bullet$ & & & & & \\ \hline
{12} & & & & & & & $\circ$ & & & & $\circ$ & \cellcolor{gray} & $\bullet$ & & & & \\ \hline
{13} & & & $\circ$ & & & & &  & & & & $\circ$ & \cellcolor{gray} & $\bullet$ & & & \\ \hline
{14} & & & & & & &  & &  & $\circ$ &  & & $\circ$ & \cellcolor{gray} & $\bullet$ & & \\ \hline
{15} & & $\circ$ & & & & & & & & & & &  & $\circ$ & \cellcolor{gray} & $\bullet$ & \\ \hline
{16} & & & & & & $\circ$ & & &  & &  & &  & & $\circ$ & \cellcolor{gray} & $\bullet$ \\ \hline
{17} & $\circ$ & & & $\circ$ & & & & &  & & & &  & & & $\circ$ & \cellcolor{gray} \\ \hline
\end{tabular}
	\renewcommand{\baselinestretch}{1} 
	\normalsize
	\caption{The bullets in the matrix shows the edges of the graph}
	\end{table}
	
	\newpage
	
	Edges

\begin{multicols*}{2}

	1-2
	
	1-8
	
	1-11
	
	1-17
	
	2-3
 
 2-15
 
 3-4
 
 3-13
 
 4-5
 
 4-17
  
 5-6
 
 5-9
 
 6-7
 
 6-16
 
 7-8
 
 7-12
 
 8-9
  
 9-10
 
 10-11
 
 10-14
 
 11-12
 
 12-13
 
 13-14
 
 14-15
 
 15-16
 
 16-17
 
 \end{multicols*}
 
 \newpage
 
 \begin{figure}[h]
\caption{$n=18$, $k=3$, $d=3$ example: (3,3)-graph on 18 vertices}
\begin{center} \includegraphics[width=0.4\textwidth,  height=0.19\paperheight]{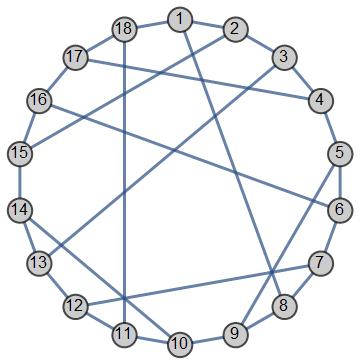}\\
'Graph6' format: \copyablespace{QhCGKCH?G?\_PG@?Cg?GG@C?E?GG} \end{center} 
\end{figure}

\begin{table}[ht!]
\centering
\renewcommand{\baselinestretch}{1.05} 
\normalsize
\begin{tabular}{|m{0.3cm}|m{0.252cm}|m{0.252cm}|m{0.252cm}|m{0.252cm}|m{0.252cm}|m{0.252cm}|m{0.252cm}|m{0.252cm}|m{0.252cm}|m{0.252cm}|m{0.252cm}|m{0.252cm}|m{0.252cm}|m{0.252cm}|m{0.252cm}|m{0.252cm}|m{0.252cm}|m{0.252cm}|m{0.252cm}|} 

\hline
\text{ }& {1} & {2} & {3} & {4} & {5} & {6} & {7} & {8} & {9} & {10}  & {11} & {12} & {13} & {14} & {15}& {16} & {17} & {18}\\ \hline
{1} & \cellcolor{gray}  & $\bullet$ &  &  & & &  & $\bullet$ & &  & & & & & & & & $\bullet$ \\ \hline
{2} & $\circ$ & \cellcolor{gray} & $\bullet$ & &  &  & & & & & & & & & $\bullet$ & & & \\ \hline
{3} & & $\circ$  & \cellcolor{gray} & $\bullet$ & & & & & & & & & $\bullet$ & & & & & \\ \hline
{4} & & & $\circ$ & \cellcolor{gray} & $\bullet$ & & & & & & & & & & & & $\bullet$ & \\ \hline
{5} & & & & $\circ$ & \cellcolor{gray} & $\bullet$ & & & $\bullet$ & & &  & & & & & & \\ \hline
{6} & & & & & $\circ$ & \cellcolor{gray} & $\bullet$ & & & & &  & & & & $\bullet$ & &  \\ \hline
{7} &  &  & & & & $\circ$ & \cellcolor{gray} & $\bullet$ & & & & $\bullet$ & & & & & & \\ \hline
{8} & $\circ$ & & & & & & $\circ$ & \cellcolor{gray} & $\bullet$ & & & & & & & & & \\ \hline
{9} & & & & & $\circ$ & & & $\circ$ & \cellcolor{gray} & $\bullet$ & & & & & & & & \\ \hline
{10} & & & & & &  & & & $\circ$ & \cellcolor{gray} & $\bullet$ & & & $\bullet$ & & & & \\ \hline
{11} & & & & & & & & & & $\circ$ & \cellcolor{gray} & $\bullet$ & & & & & & $\bullet$ \\ \hline
{12} & & & & & & & $\circ$ & & & & $\circ$ & \cellcolor{gray} & $\bullet$ & & & & & \\ \hline
{13} & & & $\circ$ & & & & &  & & & & $\circ$ & \cellcolor{gray} & $\bullet$ & & & & \\ \hline
{14} & & & & & & &  & &  & $\circ$ &  & & $\circ$ & \cellcolor{gray} & $\bullet$ & & & \\ \hline
{15} & & $\circ$ & & & & & & & & & & &  & $\circ$ & \cellcolor{gray} & $\bullet$ & & \\ \hline
{16} & & & & & & $\circ$ & & &  & &  & &  & & $\circ$ & \cellcolor{gray} & $\bullet$ & \\ \hline
{17} & & & & $\circ$ & & & & &  & & & &  & & & $\circ$ & \cellcolor{gray} & $\bullet$ \\ \hline
{18} & $\circ$ & & & & & & & &  & & $\circ$ & &  & & & & $\circ$ & \cellcolor{gray} \\ \hline
\end{tabular}
	\renewcommand{\baselinestretch}{1} 
	\normalsize
	\caption{The bullets in the matrix shows the edges of the graph}
	\end{table}
	
	\newpage
	
	Edges

\begin{multicols*}{2}

	1-2
	
	1-8
	
	1-18
	
	2-3
 
 2-15
 
 3-4
 
 3-13
 
 4-5
 
 4-17
  
 5-6
 
 5-9
 
 6-7
 
 6-16
 
 7-8
 
 7-12
 
 8-9
  
 9-10
 
 10-11
 
 10-14
 
 11-12
 
 11-18
 
 12-13
 
 13-14
 
 14-15
 
 15-16
 
 16-17
 
 17-18
 
 \end{multicols*}
 
 \newpage
 
  \begin{figure}[h]
\caption{$n=19$, $k=3$, $d=3$ example}
\begin{center} \includegraphics[width=0.4\textwidth,  height=0.19\paperheight]{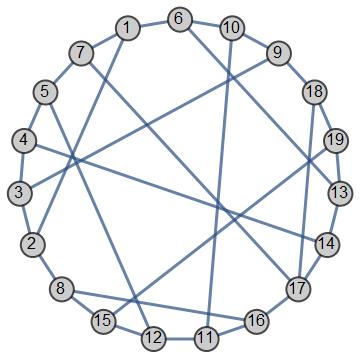}\\
'Graph6' format: \copyablespace{RhECQ?\_@G?`@@?C?\_G\_AO?\_S?\_G?DG} \end{center} 
\end{figure}

\begin{table}[ht!]
\centering
\renewcommand{\baselinestretch}{1} 
\normalsize
\begin{tabular}{|m{0.3cm}|m{0.24cm}|m{0.24cm}|m{0.24cm}|m{0.24cm}|m{0.24cm}|m{0.24cm}|m{0.24cm}|m{0.24cm}|m{0.24cm}|m{0.24cm}|m{0.24cm}|m{0.24cm}|m{0.24cm}|m{0.24cm}|m{0.24cm}|m{0.24cm}|m{0.24cm}|m{0.24cm}|m{0.24cm}|m{0.24cm}|} 

\hline
\text{ }& {1} & {2} & {3} & {4} & {5} & {6} & {7} & {8} & {9} & {10}  & {11} & {12} & {13} & {14} & {15}& {16} & {17} & {18} & {19}\\ \hline
{1} & \cellcolor{gray}  & $\bullet$ &  &  & & $\bullet$ & $\bullet$ & & &  & & & & & & & & & \\ \hline
{2} & $\circ$ & \cellcolor{gray} & $\bullet$ & &  &  & & $\bullet$ & & & & & & & & & & & \\ \hline
{3} & & $\circ$  & \cellcolor{gray} & $\bullet$ & & & & & $\bullet$ & & & & & & & & & & \\ \hline
{4} & & & $\circ$ & \cellcolor{gray} & $\bullet$ & & & & & & & & & $\bullet$ & & & & & \\ \hline
{5} & & & & $\circ$ & \cellcolor{gray} & & $\bullet$ & & & & & $\bullet$ & & & & & & & \\ \hline
{6} & $\circ$ & & & & & \cellcolor{gray} & & & & $\bullet$ & & & $\bullet$ & & & & & & \\ \hline
{7} & $\circ$ & & & & $\circ$ & & \cellcolor{gray} & & & & & & & & & & $\bullet$ & & \\ \hline
{8} & & $\circ$ & & & & & & \cellcolor{gray} & & & & & & & $\bullet$ & $\bullet$ & & & \\ \hline
{9} & & & $\circ$ & & & & & & \cellcolor{gray} & $\bullet$ & & & & & & & & $\bullet$ & \\ \hline
{10} & & & & & & $\circ$ & & & $\circ$ & \cellcolor{gray} & $\bullet$ & & & & & & & &\\ \hline
{11} & & & & & & & & & & $\circ$ & \cellcolor{gray} & $\bullet$ & & & & $\bullet$ & & & \\ \hline
{12} & & & & & $\circ$ & &  & & & & $\circ$ & \cellcolor{gray} & & & $\bullet$ & & & & \\ \hline
{13} & & & & & & $\circ$ & &  & & & & & \cellcolor{gray} & $\bullet$ & & & & & $\bullet$ \\ \hline
{14} & & & & $\circ$ & & &  & &  & &  & & $\circ$ & \cellcolor{gray} & & & $\bullet$ & & \\ \hline
{15} & & & & & & & & $\circ$ & & & & $\circ$ &  & & \cellcolor{gray} & & & & $\bullet$ \\ \hline
{16} & & & & & &  & & $\circ$ &  & & $\circ$ & &  & & & \cellcolor{gray} & $\bullet$ & & \\ \hline
{17} & & & &  & & & $\circ$ & &  & & & &  & $\circ$ & & $\circ$ & \cellcolor{gray} & $\bullet$ & \\ \hline
{18} & & & & & & & & & $\circ$ & & & &  & & & & $\circ$ & \cellcolor{gray} & $\bullet$ \\ \hline
{19} & & & & & & & & &  & & & & $\circ$ & & $\circ$ & & & $\circ$ & \cellcolor{gray} \\ \hline
\end{tabular}
	\renewcommand{\baselinestretch}{1} 
	\normalsize
	\caption{The bullets in the matrix shows the edges of the graph}
	\end{table}
	
	\newpage
	
	Edges

\begin{multicols*}{2}

	1-2
	
	1-6
	
	1-7
	
	2-3
 
 2-8
 
 3-4
 
 3-9
 
 4-5
 
 4-14
  
 5-7
 
 5-12
 
 6-10
 
 6-13
 
 7-17
 
 8-15
 
 8-16
  
 9-10
 
 9-18
 
 10-11
 
 11-12
 
 11-16
 
 12-15
 
 13-14
 
 13-19
 
 14-17
 
 15-19
 
 16-17
 
 17-18
 
 18-19
 
 \end{multicols*}
 
 \newpage

  \begin{figure}[h]
\caption{$n=20$, $k=3$, $d=3$ example: (3,3)-graph on 20 vertices (C5xF4)}
\begin{center} \includegraphics[width=0.4\textwidth,  height=0.19\paperheight]{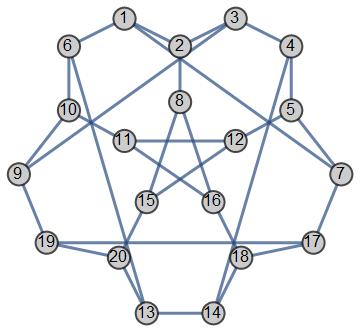}\\
\small
'Graph6' format: \copyablespace{ShECQ?\_@G?`@@?C?\_G\_AO?\_??@W@?O?DC} \end{center} 
\end{figure}

\begin{table}[ht!]
\centering
\renewcommand{\baselinestretch}{0.95} 
\normalsize
\begin{tabular}{|m{0.3cm}|m{0.228cm}|m{0.228cm}|m{0.228cm}|m{0.228cm}|m{0.228cm}|m{0.228cm}|m{0.228cm}|m{0.228cm}|m{0.228cm}|m{0.228cm}|m{0.228cm}|m{0.228cm}|m{0.228cm}|m{0.228cm}|m{0.228cm}|m{0.228cm}|m{0.228cm}|m{0.228cm}|m{0.228cm}|m{0.228cm}|m{0.228cm}|} 

\hline
\text{ }& {1} & {2} & {3} & {4} & {5} & {6} & {7} & {8} & {9} & {10}  & {11} & {12} & {13} & {14} & {15}& {16} & {17} & {18} & {19}& {20}\\ \hline
{1} & \cellcolor{gray}  & $\bullet$ &  &  & & $\bullet$ & $\bullet$ & & &  & & & & & & & & & & \\ \hline
{2} & $\circ$ & \cellcolor{gray} & $\bullet$ & &  &  & & $\bullet$ & & & & & & & & & & & & \\ \hline
{3} & & $\circ$  & \cellcolor{gray} & $\bullet$ & & & & & $\bullet$ & & & & & & & & & & & \\ \hline
{4} & & & $\circ$ & \cellcolor{gray} & $\bullet$ & & & & & & & & & $\bullet$ & & & & & & \\ \hline
{5} & & & & $\circ$ & \cellcolor{gray} & & $\bullet$ & & & & & $\bullet$ & & & & & & & & \\ \hline
{6} & $\circ$ & & & & & \cellcolor{gray} & & & & $\bullet$ & & & $\bullet$ & & & & & & & \\ \hline
{7} & $\circ$ & & & & $\circ$ & & \cellcolor{gray} & & & & & & & & & & $\bullet$ & & & \\ \hline
{8} & & $\circ$ & & & & & & \cellcolor{gray} & & & & & & & $\bullet$ & $\bullet$ & & & & \\ \hline
{9} & & & $\circ$ & & & & & & \cellcolor{gray} & $\bullet$ & & & & & & & & & $\bullet$ & \\ \hline
{10} & & & & & & $\circ$ & & & $\circ$ & \cellcolor{gray} & $\bullet$ & & & & & & & & & \\ \hline
{11} & & & & & & & & & & $\circ$ & \cellcolor{gray} & $\bullet$ & & & & $\bullet$ & & & & \\ \hline
{12} & & & & & $\circ$ & &  & & & & $\circ$ & \cellcolor{gray} & & & $\bullet$ & & & & & \\ \hline
{13} & & & & & & $\circ$ & &  & & & & & \cellcolor{gray} & $\bullet$ & & & & & & $\bullet$ \\ \hline
{14} & & & & $\circ$ & & &  & &  & &  & & $\circ$ & \cellcolor{gray} & & & & $\bullet$ & & \\ \hline
{15} & & & & & & & & $\circ$ & & & & $\circ$ &  & & \cellcolor{gray} & & & & & $\bullet$ \\ \hline
{16} & & & & & &  & & $\circ$ &  & & $\circ$ & &  & & & \cellcolor{gray} & & $\bullet$ & & \\ \hline
{17} & & & &  & & & $\circ$ & &  & & & &  & & & & \cellcolor{gray} & $\bullet$ & $\bullet$ & \\ \hline
{18} & & & & & & & & & & & & &  & $\circ$ & & $\circ$ & $\circ$ & \cellcolor{gray} & & \\ \hline
{19} & & & & & & & & & $\circ$ & & & & & &  & & $\circ$ & & \cellcolor{gray} & $\bullet$\\ \hline
{20} & & & & & & & & &  & & & & $\circ$ & & $\circ$ & & & & $\circ$ & \cellcolor{gray} \\ \hline
\end{tabular}
	\renewcommand{\baselinestretch}{1} 
	\normalsize
	\caption{The bullets in the matrix shows the edges of the graph}
	\end{table}
	
	\newpage
	
	Edges

\begin{multicols*}{2}

	1-2
	
	1-6
	
	1-7
	
	2-3
 
 2-8
 
 3-4
 
 3-9
 
 4-5
 
 4-14
  
 5-7
 
 5-12
 
 6-10
 
 6-13
 
 7-17
 
 8-15
 
 8-16
  
 9-10
 
 9-19
 
 10-11
 
 11-12
 
 11-16
 
 12-15
 
 13-14
 
 13-20
 
 14-18
 
 15-20
 
 16-18
 
 17-18
 
 17-19
 
 19-20
 
 \end{multicols*}
 
 \newpage
\section{Appendix3: $k=4$ and $d=2$ graphs}

\begin{figure}[h]
\caption{$n=11$, $k=4$, $d=2$ example: 4-Andr\'asfai graph}
\begin{center} \includegraphics[width=0.4\textwidth,  height=0.2\paperheight]{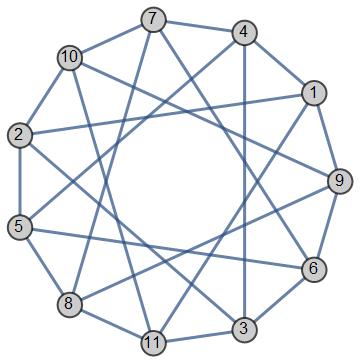}\\
'Graph6' format: \copyablespace{JlSggUDOlA\_} \end{center} 
\end{figure}

\begin{table}[ht!]
\centering
\renewcommand{\baselinestretch}{1.25} 
\normalsize
\begin{tabular}{|m{0.4cm}|m{0.3cm}|m{0.3cm}|m{0.3cm}|m{0.3cm}|m{0.3cm}|m{0.3cm}|m{0.3cm}|m{0.3cm}|m{0.3cm}|m{0.3cm}|m{0.3cm}|} 

\hline
\text{ }& {1} & {2} & {3} & {4} & {5} & {6} & {7} & {8} & {9} & {10}  & {11}\\ \hline
{1} & \cellcolor{gray}  & $\bullet$ & & $\bullet$ & & &  & & $\bullet$ & & $\bullet$ \\ \hline
{2} & $\circ$ & \cellcolor{gray} & $\bullet$ & & $\bullet$ & & & & & $\bullet$ &\\ \hline
{3} & & $\circ$ & \cellcolor{gray} & $\bullet$ & & $\bullet$ & & & & & $\bullet$ \\ \hline
{4} & $\circ$ & & $\circ$ & \cellcolor{gray} & $\bullet$ & & $\bullet$ & & & & \\ \hline
{5} & & $\circ$ & & $\circ$ & \cellcolor{gray} & $\bullet$ & & $\bullet$ & & & \\ \hline
{6} & & & $\circ$ & & $\circ$ & \cellcolor{gray} & $\bullet$ & & $\bullet$ & & \\ \hline
{7} & & & & $\circ$ & & $\circ$ & \cellcolor{gray} & $\bullet$ & & $\bullet$ & \\ \hline
{8} & &  & & & $\circ$ & & $\circ$ & \cellcolor{gray} & $\bullet$ & & $\bullet$ \\ \hline
{9} & $\circ$ & & & & & $\circ$ & & $\circ$ & \cellcolor{gray} & $\bullet$ & \\ \hline
{10} & & $\circ$ & & & &  & $\circ$ & & $\circ$ & \cellcolor{gray} & $\bullet$\\ \hline
{11} & $\circ$ & & $\circ$ & & & & &  $\circ$  & & $\circ$ & \cellcolor{gray}\\ \hline

\end{tabular}
	\renewcommand{\baselinestretch}{1} 
	\normalsize
	\caption{The bullets in the matrix shows the edges of the graph}
	\end{table}
	
\newpage
	
	Edges

\begin{multicols*}{2}
	
	1-2
	
	1-4
	
	1-9
	
 1-11
 
 2-3

 2-5
 
 2-10
 
 3-4
  
 3-16

 3-11
 
 4-5

 4-7
 
 5-6
 
 5-8
 
 6-7
 
  6-9
  
  7-8
  
 7-10
 
 8-9
 
 8-11
 
 9-10
 
 10-11
 
 \end{multicols*}

\newpage

\newpage
\begin{figure}[h]
\caption{$n=12$, $k=4$, $d=2$ example: Chv\'atal graph}
\begin{center} \includegraphics[width=0.4\textwidth,  height=0.2\paperheight]{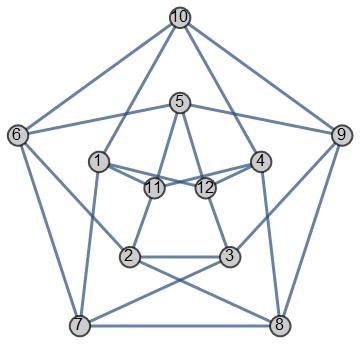}\\
'Graph6' format: \copyablespace{KG@LIchdMoV?} \end{center} 
\end{figure}

\begin{table}[ht!]
\centering
\renewcommand{\baselinestretch}{1.25} 
\normalsize
\begin{tabular}{|m{0.4cm}|m{0.3cm}|m{0.3cm}|m{0.3cm}|m{0.3cm}|m{0.3cm}|m{0.3cm}|m{0.3cm}|m{0.3cm}|m{0.3cm}|m{0.3cm}|m{0.3cm}|m{0.3cm}|} 

\hline
\text{ }& {1} & {2} & {3} & {4} & {5} & {6} & {7} & {8} & {9} & {10}  & {11} & {12}\\ \hline
{1} & \cellcolor{gray}  &  & &  & & & $\bullet$ & & & $\bullet$ & $\bullet$ & $\bullet$ \\ \hline
{2} & & \cellcolor{gray} & $\bullet$ & &  & $\bullet$ & & $\bullet$ & & & $\bullet$ &\\ \hline
{3} & & $\circ$ & \cellcolor{gray} & & & & $\bullet$ & & $\bullet$ & & & $\bullet$ \\ \hline
{4} & & & & \cellcolor{gray} &  & & & $\bullet$ & & $\bullet$ & $\bullet$ & $\bullet$ \\ \hline
{5} & & & & & \cellcolor{gray} & $\bullet$ & & & $\bullet$ & & $\bullet$ & $\bullet$ \\ \hline
{6} & & $\circ$ & & & $\circ$ & \cellcolor{gray} & $\bullet$ & & & $\bullet$ & & \\ \hline
{7} & $\circ$ & & $\circ$ & & & $\circ$ & \cellcolor{gray} & $\bullet$ & & & & \\ \hline
{8} & & $\circ$  & & $\circ$ & & & $\circ$ & \cellcolor{gray} & $\bullet$ & & & \\ \hline
{9} & & & $\circ$ & & $\circ$ & & & $\circ$ & \cellcolor{gray} & $\bullet$ & & \\ \hline
{10} & $\circ$ & & & $\circ$ & & $\circ$ & & & $\circ$ & \cellcolor{gray} & & \\ \hline
{11} & $\circ$ & $\circ$ & & $\circ$ & $\circ$ & & & & & & \cellcolor{gray} & \\ \hline
{12} & $\circ$ & & $\circ$ & $\circ$ & $\circ$ & & &  & & & & \cellcolor{gray}\\ \hline

\end{tabular}
	\renewcommand{\baselinestretch}{1} 
	\normalsize
	\caption{The bullets in the matrix shows the edges of the graph}
	\end{table}
	
	\newpage
	
	Edges

\begin{multicols*}{2}

	1-7
	
	1-10
	
	1-11
	
	1-12
 
 2-3

 2-6
 
 2-8
 
 2-11
 
 3-7
  
 3-9
 
 3-12
 
 4-8
 
 4-10
 
 4-11
 
 4-12
 
 5-6
 
 5-9
 
 5-11
 
 5-12
 
 6-7
 
 6-10
 
 7-8
  
 8-9
 
 9-10
 
 \end{multicols*}

\newpage

\begin{figure}[h]
\caption{$n=13$, $k=4$, $d=2$ example: 13-cyclotomic graph}
\begin{center} \includegraphics[width=0.4\textwidth,  height=0.2\paperheight]{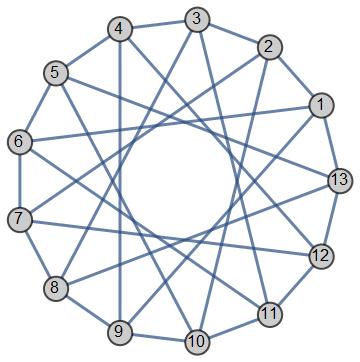}\\
'Graph6' format: \copyablespace{LhEIHEPQHGaPaP} \end{center} 
\end{figure}

\begin{table}[ht!]
\centering
\renewcommand{\baselinestretch}{1.25} 
\normalsize
\begin{tabular}{|m{0.4cm}|m{0.3cm}|m{0.3cm}|m{0.3cm}|m{0.3cm}|m{0.3cm}|m{0.3cm}|m{0.3cm}|m{0.3cm}|m{0.3cm}|m{0.3cm}|m{0.3cm}|m{0.3cm}|m{0.3cm}|} 

\hline
\text{ }& {1} & {2} & {3} & {4} & {5} & {6} & {7} & {8} & {9} & {10}  & {11} & {12} & {13}\\ \hline
{1} & \cellcolor{gray}  & $\bullet$ & &  &  & $\bullet$ &  & & $\bullet$ & & & & $\bullet$ \\ \hline
{2} & $\circ$ & \cellcolor{gray} & $\bullet$ & &  &  & $\bullet$ & & & $\bullet$ & & &\\ \hline
{3} & & $\circ$ & \cellcolor{gray} & $\bullet$ & & & & $\bullet$ & & & $\bullet$ & & \\ \hline
{4} & & & $\circ$ & \cellcolor{gray} & $\bullet$ & & & & $\bullet$ & & & $\bullet$ & \\ \hline
{5} & & & & $\circ$ & \cellcolor{gray} & $\bullet$ & & & & $\bullet$ & & & $\bullet$ \\ \hline
{6} & $\circ$ & & & & $\circ$ & \cellcolor{gray} & $\bullet$ & & & & $\bullet$ & & \\ \hline
{7} & & $\circ$ & & & & $\circ$ & \cellcolor{gray} & $\bullet$ & & & & $\bullet$ & \\ \hline
{8} & &  & $\circ$ & & & & $\circ$ & \cellcolor{gray} & $\bullet$ & & & & $\bullet$ \\ \hline
{9} & $\circ$ & & & $\circ$ & & & & $\circ$ & \cellcolor{gray} & $\bullet$ & & & \\ \hline
{10} & & $\circ$ & & & $\circ$ &  & & & $\circ$ & \cellcolor{gray} & $\bullet$ & & \\ \hline
{11} & & & $\circ$ & & & $\circ$ & & & & $\circ$ & \cellcolor{gray} & $\bullet$ &\\ \hline
{12} & & & & $\circ$ & & & $\circ$ &  & & & $\circ$ & \cellcolor{gray} & $\bullet$ \\ \hline
{13} & $\circ$ & & & & $\circ$ & & & $\circ$  & & & & $\circ$ & \cellcolor{gray} \\ \hline

\end{tabular}
	\renewcommand{\baselinestretch}{1} 
	\normalsize
	\caption{The bullets in the matrix shows the edges of the graph}
	\end{table}
	
	\newpage
	
	Edges

\begin{multicols*}{2}

	1-2
	
	1-6
	
	1-9
	
	1-13
 
 2-3

 2-7
 
 2-10
 
 3-4
  
 3-8
 
 3-11
 
 4-5
 
 4-9
 
 4-12
 
 5-6
 
 5-10
 
 5-13
 
 6-7
 
 6-11
 
 7-8
 
 7-12
  
 8-9
 
 8-13
 
 9-10
 
 10-11
 
 11-12
 
 12-13
 
 \end{multicols*}
 
 \newpage

 \begin{figure}[h]
\caption{$n=14$, $k=4$, $d=2$ example}
\begin{center} \includegraphics[width=0.4\textwidth,  height=0.2\paperheight]{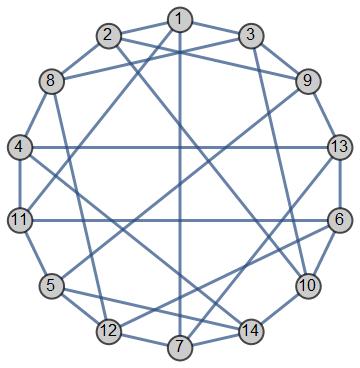}\\
'Graph6' format: \copyablespace{Mo?CB`gXCw@wDgEc?} \end{center} 
\end{figure}

\begin{table}[ht!]
\centering
\renewcommand{\baselinestretch}{1.25} 
\normalsize
\begin{tabular}{|m{0.4cm}|m{0.3cm}|m{0.3cm}|m{0.3cm}|m{0.3cm}|m{0.3cm}|m{0.3cm}|m{0.3cm}|m{0.3cm}|m{0.3cm}|m{0.3cm}|m{0.3cm}|m{0.3cm}|m{0.3cm}|m{0.3cm}|} 

\hline
\text{ }& {1} & {2} & {3} & {4} & {5} & {6} & {7} & {8} & {9} & {10}  & {11} & {12} & {13} & {14}\\ \hline
{1} & \cellcolor{gray}  & $\bullet$ & $\bullet$ &  & & & $\bullet$ & & &  & $\bullet$ & & & \\ \hline
{2} & $\circ$ & \cellcolor{gray} & & &  &  & & $\bullet$ & $\bullet$ & $\bullet$ &  & & &\\ \hline
{3} & $\circ$ & & \cellcolor{gray} & & & & & $\bullet$ & $\bullet$ & $\bullet$ & & & & \\ \hline
{4} & & & & \cellcolor{gray} & & & & $\bullet$ & & & $\bullet$ & & $\bullet$ & $\bullet$ \\ \hline
{5} & & & & & \cellcolor{gray} & & & & $\bullet$ & & $\bullet$ & $\bullet$ & & $\bullet$ \\ \hline
{6} & & & & & & \cellcolor{gray} & & & & $\bullet$ & $\bullet$ & $\bullet$ & $\bullet$ & \\ \hline
{7} & $\circ$ &  & & & & & \cellcolor{gray} & & & & & $\bullet$ & $\bullet$ & $\bullet$ \\ \hline
{8} & & $\circ$ & $\circ$ & $\circ$ & & & & \cellcolor{gray} & & & & $\bullet$ & & \\ \hline
{9} & & $\circ$ & $\circ$ & & $\circ$ & & & & \cellcolor{gray} & & & & $\bullet$ & \\ \hline
{10} & & $\circ$ & $\circ$ & & & $\circ$ & & & & \cellcolor{gray} & & & & $\bullet$ \\ \hline
{11} & $\circ$ & & & $\circ$ & $\circ$ & $\circ$ & & & & & \cellcolor{gray} & & &\\ \hline
{12} & & & & & $\circ$ & $\circ$ & $\circ$ & $\circ$ & & & & \cellcolor{gray} & & \\ \hline
{13} & & & & $\circ$ & & $\circ$ & $\circ$ &  & $\circ$ & & & & \cellcolor{gray} & \\ \hline
{14} & & & & $\circ$ & $\circ$ & & $\circ$ & & & $\circ$ & & & & \cellcolor{gray} \\ \hline

\end{tabular}
	\renewcommand{\baselinestretch}{1} 
	\normalsize
	\caption{The bullets in the matrix shows the edges of the graph}
	\end{table}
	
	\newpage
	
	Edges

\begin{multicols*}{2}

	1-2
	
	1-3
	
	1-7
	
	1-11
	
	2-8
 
 2-9
 
 2-10
 
 3-8
 
 3-9
 
 3-10
  
 4-8
 
 4-11
 
 4-13
 
 4-14
 
 5-9
 
 5-11
 
 5-12
 
 5-14
 
 6-10
 
 6-11
 
 6-12
 
 6-13
 
 7-12
 
 7-13
 
 7-14
  
 8-12
 
 9-13
 
 10-14
 
 \end{multicols*}

\newpage

\begin{figure}[h]
\caption{$n=15$, $k=4$, $d=2$ example}
\begin{center} \includegraphics[width=0.4\textwidth,  height=0.19\paperheight]{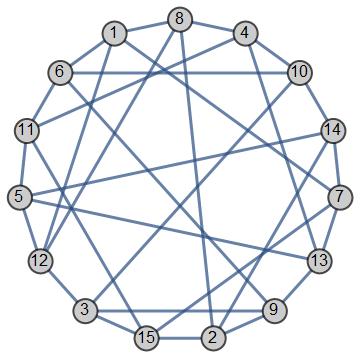}\\
'Graph6' format: \copyablespace{N?ACE`cL?wTGEgQcKP?} \end{center} 
\end{figure}

\begin{table}[ht!]
\centering
\renewcommand{\baselinestretch}{1.25} 
\normalsize
\begin{tabular}{|m{0.4cm}|m{0.3cm}|m{0.3cm}|m{0.3cm}|m{0.3cm}|m{0.3cm}|m{0.3cm}|m{0.3cm}|m{0.3cm}|m{0.3cm}|m{0.3cm}|m{0.3cm}|m{0.3cm}|m{0.3cm}|m{0.3cm}|m{0.3cm}|} 

\hline
\text{ }& {1} & {2} & {3} & {4} & {5} & {6} & {7} & {8} & {9} & {10}  & {11} & {12} & {13} & {14} & {15}\\ \hline
{1} & \cellcolor{gray}  & & & & & $\bullet$ & $\bullet$ & $\bullet$ & &  & & $\bullet$ & & & \\ \hline
{2} & & \cellcolor{gray} & & & &  & & $\bullet$ & $\bullet$ & & & & & $\bullet$ & $\bullet$\\ \hline
{3} & & & \cellcolor{gray} &  & & & & & $\bullet$ & $\bullet$ & & $\bullet$ & & & $\bullet$\\ \hline
{4} & & & & \cellcolor{gray} &  & & & $\bullet$ & & $\bullet$ & $\bullet$ & & $\bullet$ & &\\ \hline
{5} & & & & & \cellcolor{gray} & & & &  & & $\bullet$ & $\bullet$ & $\bullet$ & $\bullet$ &\\ \hline
{6} & $\circ$ & & & & & \cellcolor{gray} & & & $\bullet$ & $\bullet$ & $\bullet$ & & & & \\ \hline
{7} & $\circ$ &  & & & & & \cellcolor{gray} & & & & & & $\bullet$ & $\bullet$ & $\bullet$ \\ \hline
{8} & $\circ$ & $\circ$ &  & $\circ$ & & & & \cellcolor{gray} & & & & $\bullet$ & & & \\ \hline
{9} & & $\circ$ & $\circ$ & & & $\circ$ & & & \cellcolor{gray} & & & & $\bullet$ & & \\ \hline
{10} & & & $\circ$ & $\circ$ & & $\circ$ & & & & \cellcolor{gray} & & & & $\bullet$ & \\ \hline
{11} & & & & $\circ$ & $\circ$ & $\circ$ & & & & & \cellcolor{gray} & & & & $\bullet$ \\ \hline
{12} & $\circ$ & & $\circ$ & & $\circ$ & & & $\circ$ & & & & \cellcolor{gray} & & &  \\ \hline
{13} & & & & $\circ$ & $\circ$ & & $\circ$ &  & $\circ$ & & & & \cellcolor{gray} & & \\ \hline
{14} & & $\circ$ & & & $\circ$ & & $\circ$ & &  & $\circ$ &  & & & \cellcolor{gray} & \\ \hline
{15} & & $\circ$ & $\circ$ & & & & $\circ$ & &  & & $\circ$ & & & & \cellcolor{gray} \\ \hline

\end{tabular}
	\renewcommand{\baselinestretch}{1} 
	\normalsize
	\caption{The bullets in the matrix shows the edges of the graph}
	\end{table}
	
	\newpage
	
	Edges

\begin{multicols*}{2}

	1-6
	
	1-7
	
	1-8
	
	1-12
	
	2-8
 
 2-9
 
 2-14
 
 2-15
 
 3-9
 
 3-10
 
 3-12
 
 3-15
 
 4-8
 
 4-10
 
 4-11
 
 4-13
  
 5-11
 
 5-12
 
 5-13
 
 5-14
 
 6-9
 
 6-10
 
 6-11
 
 7-13
 
 7-14
 
 7-15
 
 8-12
  
 9-13
 
 10-14

 11-15
 
 \end{multicols*}
 
 \newpage

\section{Appendix4: $k=5$ and $d=2$ graphs}

\begin{figure}[h]
\caption{$n=16$, $k=5$, $d=2$ example: Clebsch graph}
\begin{center} \includegraphics[width=0.4\textwidth,  height=0.19\paperheight]{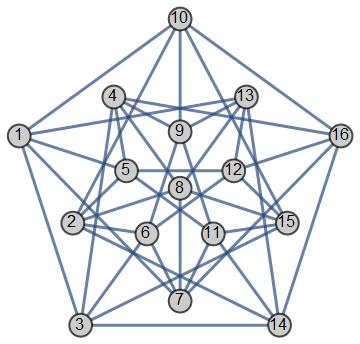}\\
'Graph6' format: \copyablespace{OPtcIcSoGT@\_\_XWAcJ\_ci} \end{center} 
\end{figure}

\begin{table}[ht!]
\centering
\renewcommand{\baselinestretch}{1.05} 
\normalsize
\begin{tabular}{|m{0.3cm}|m{0.252cm}|m{0.252cm}|m{0.252cm}|m{0.252cm}|m{0.252cm}|m{0.252cm}|m{0.252cm}|m{0.252cm}|m{0.252cm}|m{0.252cm}|m{0.252cm}|m{0.252cm}|m{0.252cm}|m{0.252cm}|m{0.252cm}|m{0.252cm}|} 

\hline
\text{ }& {1} & {2} & {3} & {4} & {5} & {6} & {7} & {8} & {9} & {10}  & {11} & {12} & {13} & {14} & {15}& {16}\\ \hline
{1} & \cellcolor{gray}  & & $\bullet$ & & $\bullet$ & & $\bullet$ & & & $\bullet$ & & & $\bullet$ & & & \\ \hline
{2} & & \cellcolor{gray} & & & $\bullet$ & $\bullet$ & & $\bullet$ & & $\bullet$ & & & & $\bullet$ & & \\ \hline
{3} & $\circ$ & & \cellcolor{gray} & $\bullet$ & & $\bullet$ & & & & & & & & $\bullet$ & $\bullet$ &\\ \hline
{4} & & & $\circ$ & \cellcolor{gray} & $\bullet$ & & & $\bullet$ & $\bullet$ & & & & & & & $\bullet$ \\ \hline
{5} & $\circ$ & $\circ$ & & $\circ$ & \cellcolor{gray} & & & & & & $\bullet$ & $\bullet$ & & & &\\ \hline
{6} & & $\circ$ & $\circ$ & & & \cellcolor{gray} & $\bullet$ & & $\bullet$ & & & $\bullet$ & & & & \\ \hline
{7} & $\circ$ &  & & & & $\circ$ & \cellcolor{gray} & $\bullet$ & & & $\bullet$ & & & & & $\bullet$ \\ \hline
{8} & & $\circ$ & & $\circ$ & & & $\circ$ & \cellcolor{gray} & & & & & $\bullet$ & & $\bullet$ & \\ \hline
{9} & & & & $\circ$ & & $\circ$ & & & \cellcolor{gray} & $\bullet$ & $\bullet$ & & $\bullet$ & & & \\ \hline
{10} & $\circ$ & $\circ$ & & & &  & & & $\circ$ & \cellcolor{gray} & & & & & $\bullet$ & $\bullet$ \\ \hline
{11} & & & & & $\circ$ & & $\circ$ & & $\circ$ & & \cellcolor{gray} & & & $\bullet$ & $\bullet$ & \\ \hline
{12} & & & & & $\circ$ & $\circ$ & & & & & & \cellcolor{gray} & $\bullet$ & & $\bullet$ & $\bullet$ \\ \hline
{13} & $\circ$ & & & & & & & $\circ$ & $\circ$ & & & $\circ$ & \cellcolor{gray} & $\bullet$ & & \\ \hline
{14} & & $\circ$ & $\circ$ & & & & & &  & & $\circ$ & & $\circ$ & \cellcolor{gray} & & $\bullet$ \\ \hline
{15} & & & $\circ$ & & & & & $\circ$ & & $\circ$ & $\circ$ & $\circ$ &  & & \cellcolor{gray} & \\ \hline
{16} & & & & $\circ$ & & & $\circ$ & &  & $\circ$ & & $\circ$ & & $\circ$ & & \cellcolor{gray} \\ \hline

\end{tabular}
	\renewcommand{\baselinestretch}{1} 
	\normalsize
	\caption{The bullets in the matrix shows the edges of the graph}
	\end{table}
	
	\newpage
	
	Edges

\begin{multicols*}{2}

	1-3
	
	1-5
	
	1-7
	
	1-10
	
	1-13
	
	2-5
 
 2-6
 
 2-8
 
 2-10
 
 2-14
 
 3-4
 
 3-6
 
 3-14
 
 3-15
 
 4-5
 
 4-8
 
 4-9
 
 4-16
  
 5-11
 
 5-12
 
 6-7
 
 6-9
 
 6-12
 
 7-8
 
 7-11
 
 7-16
 
 8-13
  
 8-15
 
 9-10
 
 9-11
 
 9-13
 
 10-15
 
 10-16
 
 11-14
 
 11-15
 
 12-13
 
 12-15
 
 12-16
 
 13-14
 
 14-16
 
 \end{multicols*}
 
\newpage

\begin{figure}[h]
\caption{$n=17$, $k=5$, $d=2$ example}
\begin{center} \includegraphics[width=0.4\textwidth,  height=0.19\paperheight]{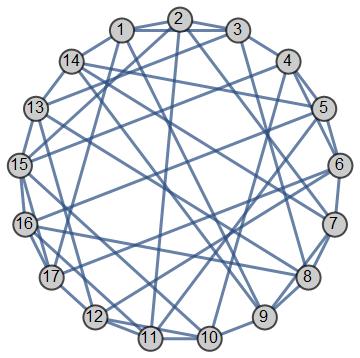}\\
'Graph6' format: \copyablespace{PxCYHEBCIO\_bGPagiAOQP`@K} \end{center} 
\end{figure}

\begin{table}[ht!]
\centering
\renewcommand{\baselinestretch}{1.05} 
\normalsize
\begin{tabular}{|m{0.3cm}|m{0.252cm}|m{0.252cm}|m{0.252cm}|m{0.252cm}|m{0.252cm}|m{0.252cm}|m{0.252cm}|m{0.252cm}|m{0.252cm}|m{0.252cm}|m{0.252cm}|m{0.252cm}|m{0.252cm}|m{0.252cm}|m{0.252cm}|m{0.252cm}|m{0.252cm}|} 

\hline
\text{ }& {1} & {2} & {3} & {4} & {5} & {6} & {7} & {8} & {9} & {10}  & {11} & {12} & {13} & {14} & {15}& {16} & {17}\\ \hline
{1} & \cellcolor{gray}  & $\bullet$ & $\bullet$ &  & & &  & & $\bullet$ &  &  & & & $\bullet$ & & & $\bullet$ \\ \hline
{2} & $\circ$ & \cellcolor{gray} & $\bullet$ & &  &  & $\bullet$ & & & & $\bullet$ & & & & $\bullet$ & & \\ \hline
{3} & $\circ$ & $\circ$  & \cellcolor{gray} & $\bullet$ & & & & $\bullet$ & & & & & $\bullet$ & & & & \\ \hline
{4} & & & $\circ$ & \cellcolor{gray} & $\bullet$ & $\bullet$ & & & & $\bullet$ & & & & & $\bullet$ & & \\ \hline
{5} & & & & $\circ$ & \cellcolor{gray} & $\bullet$ & & & & & $\bullet$ &  & & $\bullet$ & & $\bullet$ & \\ \hline
{6} & & & & $\circ$ & $\circ$ & \cellcolor{gray} & $\bullet$ & & & & & $\bullet$ & & & & & $\bullet$ \\ \hline
{7} &  & $\circ$ & & & & $\circ$ & \cellcolor{gray} & $\bullet$ & $\bullet$ & & & & & $\bullet$ & & & \\ \hline
{8} & & & $\circ$ & & & & $\circ$ & \cellcolor{gray} & $\bullet$ & & & & $\bullet$ & & & $\bullet$ & \\ \hline
{9} & $\circ$ & & & & & & $\circ$ & $\circ$ & \cellcolor{gray} & $\bullet$ & & & & $\bullet$ & & & \\ \hline
{10} & & & & $\circ$ & &  & & & $\circ$ & \cellcolor{gray} & $\bullet$ & $\bullet$ & & & $\bullet$ & & \\ \hline
{11} & & $\circ$ & & & $\circ$ & & & & & $\circ$ & \cellcolor{gray} & $\bullet$ & & & & $\bullet$ & \\ \hline
{12} & & & & & & $\circ$ & & & & $\circ$ & $\circ$ & \cellcolor{gray} & $\bullet$ & & & & $\bullet$ \\ \hline
{13} & & & $\circ$ & & & & & $\circ$ & & & & $\circ$ & \cellcolor{gray} & $\bullet$ & $\bullet$ & & \\ \hline
{14} & $\circ$ & & & & $\circ$ & & $\circ$ & & $\circ$ & &  & & $\circ$ & \cellcolor{gray} & & & \\ \hline
{15} & & $\circ$ & & $\circ$ & & & & & & $\circ$ & & & $\circ$ & & \cellcolor{gray} & $\bullet$ & $\bullet$ \\ \hline
{16} & & & & & $\circ$ & & & $\circ$ &  & & $\circ$ & &  & & $\circ$ & \cellcolor{gray} & $\bullet$ \\ \hline
{17} & $\circ$ & & & & & $\circ$ & & &  & & & $\circ$ &  & & $\circ$ & $\circ$ & \cellcolor{gray} \\ \hline
\end{tabular}
	\renewcommand{\baselinestretch}{1} 
	\normalsize
	\caption{The bullets in the matrix shows the edges of the graph}
	\end{table}
	
	\newpage
	
	Edges

\begin{multicols*}{2}

	1-2
	
	1-3
	
	1-9
	
	1-14
	
	1-17
	
	2-3
 
 2-7
 
 2-11
 
 2-15
 
 3-4
 
 3-8
 
 3-13
 
 4-5
 
 4-6
 
 4-10
 
 4-15
  
 5-6
 
 5-11
 
 5-14
 
 5-16
 
 6-7
 
 6-12
 
 6-17
 
 7-8
 
 7-9
 
 7-14
 
 8-9
 
 8-13
 
 8-16
  
 9-10
 
 9-14
 
 10-11
 
 10-12
 
 10-15
 
 11-12
 
 11-16
 
 12-13
 
 12-17
 
 13-14
 
 13-15
 
 15-16
 
 15-17
 
 16-17
 
 \end{multicols*}
 
 \newpage
 
 \begin{figure}[h]
\caption{$n=18$, $k=5$, $d=2$ example: (18,1)-noncayley transitive graph}
\begin{center} \includegraphics[width=0.4\textwidth,  height=0.19\paperheight]{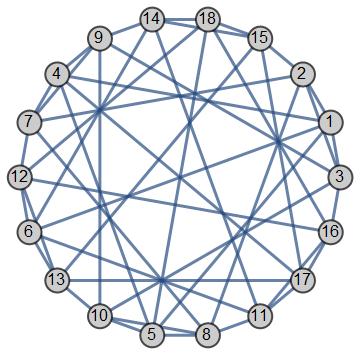}\\
'Graph6' format: \copyablespace{Q\{eAaSqIWI?o@D@IG?X?WCAkGDo} \end{center} 
\end{figure}

\begin{table}[ht!]
\centering
\renewcommand{\baselinestretch}{1.05} 
\normalsize
\begin{tabular}{|m{0.3cm}|m{0.252cm}|m{0.252cm}|m{0.252cm}|m{0.252cm}|m{0.252cm}|m{0.252cm}|m{0.252cm}|m{0.252cm}|m{0.252cm}|m{0.252cm}|m{0.252cm}|m{0.252cm}|m{0.252cm}|m{0.252cm}|m{0.252cm}|m{0.252cm}|m{0.252cm}|m{0.252cm}|} 

\hline
\text{ }& {1} & {2} & {3} & {4} & {5} & {6} & {7} & {8} & {9} & {10}  & {11} & {12} & {13} & {14} & {15}& {16} & {17} & {18}\\ \hline
{1} & \cellcolor{gray}  & $\bullet$ & $\bullet$ & $\bullet$ & $\bullet$ & $\bullet$ &  & & &  &  & & & & & & & \\ \hline
{2} & $\circ$ & \cellcolor{gray} & $\bullet$ & &  &  & $\bullet$ & $\bullet$ & & & & & & & $\bullet$ & & & \\ \hline
{3} & $\circ$ & $\circ$  & \cellcolor{gray} & & & & & & $\bullet$ & $\bullet$& & & & & & $\bullet$ & & \\ \hline
{4} & $\circ$ & & & \cellcolor{gray} & $\bullet$ & & $\bullet$ & & $\bullet$ &  & & & & &  & & $\bullet$ & \\ \hline
{5} & $\circ$ & & & $\circ$ & \cellcolor{gray} & & & $\bullet$ & & $\bullet$ & &  & & & & & & $\bullet$ \\ \hline
{6} & $\circ$ & & & & & \cellcolor{gray} &  & & & & $\bullet$ & $\bullet$ & $\bullet$ & $\bullet$ & & & & \\ \hline
{7} &  & $\circ$ & & $\circ$ & & & \cellcolor{gray} & $\bullet$ & $\bullet$ & & & $\bullet$ & & & & & & \\ \hline
{8} & & $\circ$ & & & $\circ$ & & $\circ$ & \cellcolor{gray} & & $\bullet$ & $\bullet$ & & & & & & & \\ \hline
{9} &  & & $\circ$ & $\circ$ & & & $\circ$ & & \cellcolor{gray} & $\bullet$ & & & & $\bullet$ & & & & \\ \hline
{10} & & & $\circ$ & & $\circ$ &  & & $\circ$ & $\circ$ & \cellcolor{gray} & & & $\bullet$ & & & & & \\ \hline
{11} & & & & & & $\circ$ & & $\circ$ & & & \cellcolor{gray} & & & $\bullet$ & & $\bullet$ & $\bullet$ & \\ \hline
{12} & & & & & & $\circ$ & $\circ$ & & & & & \cellcolor{gray} & $\bullet$ & & & $\bullet$ & & $\bullet$ \\ \hline
{13} & & & & & & $\circ$ & & & & $\circ$ & & $\circ$ & \cellcolor{gray} & & $\bullet$ & & $\bullet$ & \\ \hline
{14} & & & & & & $\circ$ & & & $\circ$ & & $\circ$ & & & \cellcolor{gray} & $\bullet$ & & & $\bullet$ \\ \hline
{15} & & $\circ$ & & & & & & & & & & & $\circ$ & $\circ$ & \cellcolor{gray} & & $\bullet$ & $\bullet$ \\ \hline
{16} & & & $\circ$ & & & & & &  & & $\circ$ & $\circ$ &  & & & \cellcolor{gray} & $\bullet$ & $\bullet$ \\ \hline
{17} & & & & $\circ$ & & & & &  & & $\circ$ & & $\circ$ & & $\circ$ & $\circ$ & \cellcolor{gray} & \\ \hline
{18} & & & & & $\circ$ & & & &  & & & $\circ$ &  & $\circ$ & $\circ$ & $\circ$ & & \cellcolor{gray} \\ \hline
\end{tabular}
	\renewcommand{\baselinestretch}{1} 
	\normalsize
	\caption{The bullets in the matrix shows the edges of the graph}
	\end{table}
	
	\newpage
	
	Edges

\begin{multicols*}{2}

	1-2
	
	1-3
	
	1-4
	
	1-5
	
	1-6
	
	2-3
 
 2-7
 
 2-8
 
 2-15
 
 3-9
 
 3-10
 
 3-16
 
 4-5
 
 4-7
 
 4-9
 
 4-17
  
 5-8
 
 5-10
 
 5-18
 
 6-11
 
 6-12
 
 6-13
 
 6-14
 
 7-8
 
 7-9
 
 7-12
 
 8-10
 
 8-11
  
 9-10
 
 9-14
 
 10-13
 
 11-14
 
 11-16
 
 11-17
 
 12-13
 
 12-16
 
 12-18
 
 13-15
 
 13-17
 
 14-15
 
 14-18
 
 15-17
 
 15-18
 
 16-17
 
 16-18
 
 \end{multicols*}
\newpage

\begin{figure}[h]
\caption{$n=19$, $k=5$, $d=2$ example}
\begin{center} \includegraphics[width=0.4\textwidth,  height=0.19\paperheight]{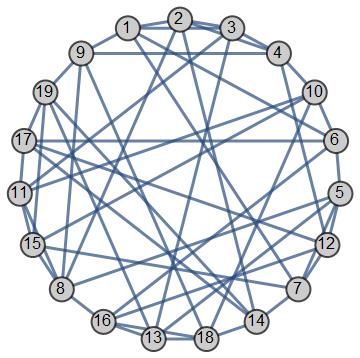}\\
'Graph6' format: \copyablespace{RzAKQQPD@AbOI?O\_?Z?IK@BO?rO@FO}  \end{center} 
\end{figure}

\begin{table}[ht!]
\centering
\renewcommand{\baselinestretch}{1} 
\normalsize
\begin{tabular}{|m{0.3cm}|m{0.24cm}|m{0.24cm}|m{0.24cm}|m{0.24cm}|m{0.24cm}|m{0.24cm}|m{0.24cm}|m{0.24cm}|m{0.24cm}|m{0.24cm}|m{0.24cm}|m{0.24cm}|m{0.24cm}|m{0.24cm}|m{0.24cm}|m{0.24cm}|m{0.24cm}|m{0.24cm}|m{0.24cm}|} 

\hline
\text{ }& {1} & {2} & {3} & {4} & {5} & {6} & {7} & {8} & {9} & {10}  & {11}& {12} & {13} & {14} & {15} & {16} & {17} & {18} & {19}\\ \hline
{1} & \cellcolor{gray}  & $\bullet$ & $\bullet$ & & & $\bullet$ & $\bullet$ & & $\bullet$ & & & & & & & & & &\\ \hline
{2} & $\circ$ & \cellcolor{gray} & $\bullet$ & $\bullet$ & & & & $\bullet$ & & & & & & $\bullet$ & & & & &\\ \hline
{3} & $\circ$ & $\circ$ & \cellcolor{gray} & $\bullet$ & & & & & & & $\bullet$ & & $\bullet$ & & & & & &\\ \hline
{4} & & $\circ$ & $\circ$ & \cellcolor{gray} & & & & & $\bullet$ & $\bullet$ & & $\bullet$ & & & & & & &\\ \hline
{5} & & & & & \cellcolor{gray} & $\bullet$ & $\bullet$ & $\bullet$ & & & & $\bullet$ & $\bullet$ & & & & & &\\ \hline
{6} & $\circ$ & & & & $\circ$ & \cellcolor{gray} & & & & $\bullet$ & & & & & & $\bullet$ & $\bullet$ & &\\ \hline
{7} & $\circ$ & & & & $\circ$ & & \cellcolor{gray} & & & & & $\bullet$ & & $\bullet$ & $\bullet$ & & & &\\ \hline
{8} & & $\circ$ & & & $\circ$ & & & \cellcolor{gray} & $\bullet$ & & $\bullet$ & & & & $\bullet$ & $\bullet$ & & &\\ \hline
{9} & $\circ$ & & & $\circ$ & & & & $\circ$ & \cellcolor{gray} & & & & & & & & & $\bullet$ & $\bullet$\\ \hline
{10} & & & & $\circ$ & & $\circ$ & & & & \cellcolor{gray} & $\bullet$ & & & & $\bullet$ & & & $\bullet$ &\\ \hline
{11} & & & $\circ$ & & & & & $\circ$ & & $\circ$ & \cellcolor{gray} & & & & $\bullet$ & & $\bullet$ & &\\ \hline
{12} & & & & $\circ$ & $\circ$ & & $\circ$ & & & & & \cellcolor{gray} & & & & $\bullet$ & $\bullet$ & &\\ \hline
{13} & & & $\circ$ & & $\circ$ & & & & & & & & \cellcolor{gray} & & & $\bullet$ & & $\bullet$ & $\bullet$ \\ \hline
{14} & & $\circ$ & & & & & $\circ$ & & & & & & & \cellcolor{gray} & & & $\bullet$ & $\bullet$ & $\bullet$ \\ \hline
{15} & & & & & & & $\circ$ & $\circ$ & & $\circ$ & $\circ$ & & & & \cellcolor{gray} & & & & $\bullet$ \\ \hline
{16} & & & & & & $\circ$ & & $\circ$ & & & & $\circ$ & $\circ$ & & & \cellcolor{gray} & & $\bullet$ &\\ \hline
{17} & & & & & & $\circ$ & & & & & $\circ$ & $\circ$ & & $\circ$ & & & \cellcolor{gray} & & $\bullet$\\ \hline
{18} & & & & & & & & & $\circ$ & $\circ$ & & & $\circ$ & $\circ$ & & $\circ$ & & \cellcolor{gray} &\\ \hline
{19} & & & & & & & & & $\circ$ & & & & $\circ$ & $\circ$ & $\circ$ & & $\circ$ & & \cellcolor{gray}\\ \hline

\end{tabular}
	\renewcommand{\baselinestretch}{1} 
	\normalsize
	\caption{The bullets in the matrix shows the edges of the graph}
	\end{table}
	
	\bigskip
	
	Edges
	
\begin{multicols*}{2}
	
	1-2
	
	1-3
	
	1-6
	
 1-7
 
 1-9
 
 2-3

 2-4
 
 2-8
 
 2-14

 3-4
  
 3-11

 3-13
 
 4-9

 4-10
 
 4-12
 
 5-6
 
 5-7
 
 5-8
 
  5-12
  
  5-13
  
 6-10
 
 6-16
 
 6-17
 
 7-12
 
 7-14
 
 7-15
 
 8-9
 
 8-15
 
 8-16
 
 8-11
 
 9-18
 
 9-19
 
 10-11
 
 10-15
 
 10-18
 
 11-15
 
 11-17
 
 12-16
 
 12-17
 
 13-16
 
 13-18
 
 13-19
 
 14-17
 
 14-18
 
 14-19
 
 15-19
 
 16-18
 
 17-19
 
 \end{multicols*}

\newpage

\begin{figure}[h]
\caption{$n=20$, $k=5$, $d=2$ example: (20,8)-noncayley transitive graph}
\begin{center} \includegraphics[width=0.4\textwidth,  height=0.19\paperheight]{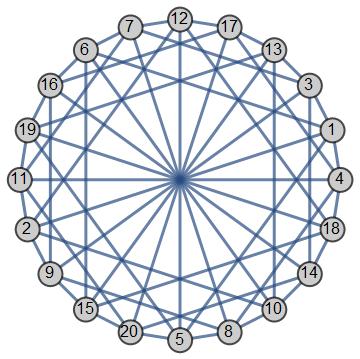}\\
\small
'Graph6' format: \copyablespace{Ssa@Gt`PQcHOGCGC?cOHAC@cOD\_OSgORO} \end{center} 
\end{figure}

\begin{table}[ht!]
\centering
\renewcommand{\baselinestretch}{0.95} 
\normalsize
\begin{tabular}{|m{0.3cm}|m{0.228cm}|m{0.228cm}|m{0.228cm}|m{0.228cm}|m{0.228cm}|m{0.228cm}|m{0.228cm}|m{0.228cm}|m{0.228cm}|m{0.228cm}|m{0.228cm}|m{0.228cm}|m{0.228cm}|m{0.228cm}|m{0.228cm}|m{0.228cm}|m{0.228cm}|m{0.228cm}|m{0.228cm}|m{0.228cm}|} 

\hline
\text{ }& {1} & {2} & {3} & {4} & {5} & {6} & {7} & {8} & {9} & {10}  & {11}& {12} & {13} & {14} & {15} & {16} & {17} & {18} & {19}  & {20}\\ \hline
{1} & \cellcolor{gray}  & $\bullet$ & $\bullet$ & $\bullet$ & $\bullet$ & $\bullet$ & & & & & & & & & & & & & & \\ \hline
{2} & $\circ$ & \cellcolor{gray} & & & & & & & $\bullet$ & $\bullet$ & $\bullet$ & $\bullet$ & & & & & & & & \\ \hline
{3} & $\circ$ & & \cellcolor{gray} & & & & $\bullet$ & & $\bullet$ & & & & $\bullet$ & $\bullet$ & & & & & & \\ \hline
{4} & $\circ$ & & & \cellcolor{gray} & & & & $\bullet$ & & & $\bullet$ & & & & & & $\bullet$ & $\bullet$ & & \\ \hline
{5} & $\circ$ & & & & \cellcolor{gray} & & & $\bullet$ & & & & $\bullet$ & & & & & & & $\bullet$ & $\bullet$ \\ \hline
{6} & $\circ$ & & & & & \cellcolor{gray} & $\bullet$ & & & $\bullet$ & & & & & $\bullet$ & $\bullet$ & & & & \\ \hline
{7} & & & $\circ$ & & & $\circ$ & \cellcolor{gray} & $\bullet$ & & & $\bullet$ & $\bullet$ & & & & & & & & \\ \hline
{8} & & & & $\circ$ & $\circ$ & & $\circ$ & \cellcolor{gray} & $\bullet$ & $\bullet$ & & & & & & & & & & \\ \hline
{9} & & $\circ$ & $\circ$ & & & & & $\circ$ & \cellcolor{gray} & & & & & & $\bullet$ & $\bullet$ & & & & \\ \hline
{10} & & $\circ$ & & & & $\circ$ & & $\circ$ & & \cellcolor{gray} & & & $\bullet$ & $\bullet$ & & & & & & \\ \hline
{11} & & $\circ$ & & $\circ$ & & & $\circ$ & & & & \cellcolor{gray} & & & & & & & & $\bullet$ & $\bullet$ \\ \hline
{12} & & $\circ$ & & & $\circ$ & & $\circ$ & & & & & \cellcolor{gray} & & & & & $\bullet$ & $\bullet$ & & \\ \hline
{13} & & & $\circ$ & & & & & & & $\circ$ & & & \cellcolor{gray} & & $\bullet$ & & $\bullet$ & & $\bullet$  & \\ \hline
{14} & & & $\circ$ & & & & & & & $\circ$ & & & & \cellcolor{gray} & & $\bullet$ & & $\bullet$ & & $\bullet$ \\ \hline
{15} & & & & & & $\circ$ & & & $\circ$ & & & & $\circ$ & & \cellcolor{gray} & & & $\bullet$ & & $\bullet$ \\ \hline
{16} & & & & & & $\circ$ & & & $\circ$ & & & & & $\circ$ & & \cellcolor{gray} & $\bullet$ & & $\bullet$ & \\ \hline
{17} & & & & $\circ$ & & & & & & & & $\circ$ & $\circ$ & & & $\circ$& \cellcolor{gray} & & & $\bullet$ \\ \hline
{18} & & & & $\circ$ & & & & & & & & $\circ$ & & $\circ$ & $\circ$ & & & \cellcolor{gray} & $\bullet$ & \\ \hline
{19} & & & & & $\circ$ & & & & & & $\circ$ & & $\circ$ & & & $\circ$ & & $\circ$ & \cellcolor{gray} & \\ \hline
{20} & & & & & $\circ$ & & & & & & $\circ$ & & & $\circ$ & $\circ$ & & $\circ$ & & & \cellcolor{gray} \\ \hline

\end{tabular}
	\renewcommand{\baselinestretch}{1} 
	\normalsize
	\caption{The bullets in the matrix shows the edges of the graph}
	\end{table}
	
	\bigskip
	
	Edges
	
\begin{multicols*}{2}
	
	1-2
	
	1-3
	
	1-4
	
 1-5
 
 1-6
 
 2-9

 2-10
 
 2-11
 
 2-12
 
 3-7
 
 3-9
  
 3-13
  
 3-14
 
 4-8
 
 4-11
 
 4-17
 
 4-18
 
 5-8
 
 5-12
 
 5-19
 
 5-20
  
 6-7
  
 6-10
 
 6-15
 
 6-16
 
 7-8
 
 7-11
 
 7-12
 
 8-9
 
 8-10
 
 9-15
 
 9-16
 
 10-13
 
 10-14
 
 11-19
 
 11-20
 
 12-17
 
 12-18
 
 13-15
 
 13-17
 
 13-19
 
 14-16
 
 14-18
 
 14-20
 
 15-18
 
 15-20
 
 16-17
 
 16-19
 
 17-20
 
 18-19
 
 \end{multicols*}

%\end{document}

\end{document}